\documentclass[a4paper, 10pt]{article}

\usepackage{draft}
\usepackage[mathscr]{eucal}
\usepackage{tikz-cd}
\usepackage{physics}
\usepackage{cancel}
\newcommand{\Base}{\mathscr{X}}
\newcommand{\Bracket}[1]{\pmb\{#1\pmb\}}
\newcommand{\Completion}{\pmb\gamma}
\newcommand{\Curvature}{\mathcal{R}}
\newcommand{\dR}{\mathrm{d}}
\newcommand{\Fedosov}{\mathfrak{D}}
\newcommand{\Fock}{\mathfrak{F}}
\newcommand{\Functions}{\mathscr{C}^\infty}
\newcommand{\R}{\mathbb{R}}
\newcommand{\Rep}[2]{#1 \pmb{\triangleright} #2}
\newcommand{\SymDer}{\partial_{\scriptscriptstyle\nabla}}
\newcommand{\Taylor}{j_{\scriptstyle\nabla}}
\newcommand{\Weyl}{\scriptstyle\mathrm{Weyl}}
\newcommand{\WeylAlg}{\mathcal{A}}
\newcommand{\WeylBundle}{\mathscr{W}}
\newcommand{\Wigner}{\mathcal{W}}

\tikzstyle{vertex}=[circle,draw,thick,fill=black,inner sep=0pt,minimum size=3pt]

\begin{document}

\title{Scalar Field on a higher-spin Background\\
    via Fedosov quantization}
\abstract{%
    Conformal higher-spin gravity is the log-divergent part
    of the effective action of the scalar field
    coupled to background fields via higher-spin currents,
    as was defined by Segal and Tseytlin,
    which can be worked out over the flat space background.
    We revisit the problem of the scalar field
    in a higher-spin background
    and propose a manifestly covariant version thereof.
    The construction utilizes the Fedosov quantization
    of the cotangent bundle and the action is written
    with the help of the trace on a curved phase space
    that is provided by the Feigin--Felder--Shoikhet cocycle. 
    The same construction allows one to formulate
    quantum mechanics on a curved space,
    the phase space being the cotangent bundle. 
}
\author{Thomas {\sc Basile}}
\author{Shailesh {\sc Dhasmana}}
\author{Evgeny {\sc Skvortsov}%
    \footnote[3]{Research Associate of the Fund
    for Scientific Research -- FNRS, Belgium}%
   \footnote[4]{Also at Lebedev Institute of Physics.}%
}
\affiliation{%
    Service de Physique de l'Univers, Champs et Gravitation, \\
    Universit\'e de Mons, 20 place du Parc, 7000 Mons, Belgium
}

\maketitle

\section{Introduction}
Conformal higher-spin gravity
\cite{Tseytlin:2002gz, Segal:2002gd, Bekaert:2010ky}
is a rare example of a higher-spin extension
of (conformal) gravity where the usual field theory concepts 
directly apply. For example, despite the infinite spectrum
of states it is a perturbatively local field theory,
a feature enforced by the Weyl symmetry;
it has an action and there is some understanding
of what the underlying geometry is.%
\footnote{Numerous $3d$ higher-spin gravities \cite{Blencowe:1988gj, Bergshoeff:1989ns, Pope:1989vj, Fradkin:1989xt, Campoleoni:2010zq, Henneaux:2010xg, Grigoriev:2019xmp, Grigoriev:2020lzu}
lack local degrees of freedom except for \cite{Sharapov:2024euk}. 
The action for chiral higher-spin gravity
\cite{Metsaev:1991mt,Metsaev:1991nb,Ponomarev:2016lrm}
is available in the light-cone gauge,
while covariant equations of motion are known 
\cite{Sharapov:2022awp,Sharapov:2022wpz,Sharapov:2022nps},
and there is no clear understanding of the geometry behind.}
The action can be extracted as the anomalous part
of the effective action of a (massless) scalar field
in a higher-spin background, which is similar to how
the Yang--Mills and the Weyl gravity actions
arise as conformal anomalies in the presence
of the background of gauge fields and of conformal gravity, 
respectively.

Despite its conceptual simplicity, e.g. the action
can be extracted order by order over the flat space background 
\cite{Bekaert:2010ky,Joung:2015eny,Bonezzi:2017mwr},
realizing conformal and higher-spin symmetries
in a manifestly covariant way has been an open problem.
This problem was solved recently in \cite{Basile:2022nou}. 
Essentially, the notion of a higher-spin covariant derivative 
is encoded in the Fedosov approach to deformation quantization 
of the cotangent bundle (of spacetime).
The notion of a higher-spin invariant measure,
which is needed to write down an action, is encoded
in the trace operation. Surprisingly, defining a trace
(over the deformed algebra of functions) \emph{explicitly}
within the Fedosov approach took some time and was done
by Feigin, Felder and Shoikhet in \cite{Feigin:2005},
who relied on Shoikhet's proof \cite{Shoikhet:2000gw}
of an extension of Kontsevich formality
conjectured by Tsygan \cite{Tsygan:1999},
known as Shoikhet--Tsygan--Kontsevich formality. 

In this paper we address the problem of how to couple
a (massless) scalar field to the background
of (off-shell) conformal higher-spin fields.
This problem was bypassed in \cite{Basile:2022nou}
thanks to Segal's approach to conformal higher-spin gravity 
that allows one to fix the action directly by
the gauge invariance and, hence, the actual problem 
\emph{boiled down} to covariantizing the construction.   

In more detail, the route between the free scalar field and conformal higher-spin fields is as follows. Being a free theory, the massless scalar on flat space
possesses an infinite tower of on-shell conserved currents
for all integer spin $s\geq0$,
which come from the invariance of the d'Alembert equation
under the action of conformal Killing tensors
\cite{Eastwood:2002su}. The existence of these conserved currents
opens the possibility of introducing interactions
between the scalar field $\phi$, and gauge fields
of arbitrary spin, starting with the Noether coupling
and completing it to all orders. For instance,
the currents of spin $1$ and $2$,
\begin{equation}
    J_\mu := \tfrac{i}2\,\big(\phi^*\,\partial_\mu\phi
    - \phi\,\partial_\mu\phi^*\big)\,,
    \qquad 
    T_{\mu\nu} := \phi^*\,\partial_\mu\partial_\nu\phi
    -\tfrac{2n}{n-1}\,\partial_{(\mu}\phi^*\partial_{\nu)}\phi
    + \phi\,\partial_\mu\partial_\nu\phi^* - \text{(traces)}\,,
\end{equation}
can be used to introduce gauge fields, say $A_\mu$
and $h_{\mu\nu}$ respectively, to the free scalar action, via
\begin{equation}
    S[\phi,A] = \int_{\R^n} \dR^nx\,\tfrac12\,\phi^*\Box\phi
    + e\,A_\mu\,J^\mu\,,
    \qquad\text{and}\qquad 
    S[\phi,h] = \int_{\R^n} \dR^nx\,\tfrac12\,\phi^*\Box\phi
    + \kappa\,T^{\mu\nu}\,h_{\mu\nu}\,,
\end{equation}
where $e$ and $\kappa$ are coupling constants.
Since both currents are divergenceless on-shell (meaning 
modulo the scalar field equation of motion $\Box\phi\approx0$),
and the spin $2$ one is also traceless,
the gauge transformations
\begin{equation}\label{eq:linear}
    \delta_\varepsilon A_\mu = \partial_\mu\varepsilon\,,
    \qquad 
    \delta_\xi h_{\mu\nu}
        = \partial_{(\mu} \xi_{\nu)} + \eta_{\mu\nu}\,\sigma\,,
\end{equation}
together with the transformations of the scalar field
\begin{equation}
    \delta_\varepsilon\phi
        = -ie\,\varepsilon\,\phi\,,
    \qquad\text{and}\qquad
    \delta_\xi \phi = \xi^\mu\partial_\mu\phi\,,
\end{equation}
leave the respective actions invariant up to second order
in the coupling constants. The spin $1$ case can be completed
to a gauge-invariant action to all orders by adding
a quadratic term in the gauge field $A_\mu$, which amounts
to re-constructing scalar electrodynamics,
\begin{equation}
    S[\phi,A] = \tfrac12\,\int_{\R^n} \dR^nx\,
    \phi^*\Box_A\phi\,,
    \qquad\text{where}\qquad
    \Box_A = (\partial_\mu + ie\,A_\mu)
    (\partial^\mu + ie\,A^\mu)\,.
\end{equation}
The spin $2$ case is technically more involved,
though similar in spirit. It requires infinitely many 
correction terms, which can be re-summed into
the action for the conformally-coupled scalar field,
\begin{equation}\label{eq:action_conformal_sc}
    S[\phi,g] = \tfrac1{2\kappa}\,\int_M \dR^nx\,\sqrt{-g}\
    \phi^*\,\big(\nabla^2 - \tfrac{n-2}{4(n-1)}\,R\big)\phi\,,
    \qquad 
    g_{\mu\nu} := \eta_{\mu\nu} + h_{\mu\nu}\,,
    \qquad 
    \nabla^2 := g^{\mu\nu}\,\nabla_\mu\nabla_\nu\,,
\end{equation}
expanded around flat spacetime. In both of these low spin cases,
the Noether coupling is completed by higher order terms
in the gauge fields and suitable deformations
of their gauge symmetries. The output of this procedure
is an action, quadratic in the scalar field,
and non-linear in the gauge fields. The all order coupling
of the former to the latter is encoded in a covariant
differential operator---the square of the covariant derivative
in the spin $1$ case and the conformal Laplacian
in the spin $2$ case. From this point of view,
these gauge fields are \emph{background fields}
for the scalar field $\phi$.

One can then integrate out the scalar field to derive 
an action for the background fields. To be more precise,
the effective action for the scalar field $\phi$
can be interpreted as an action for the background fields,
a point of view already advocated by Sakharov
\cite{Sakharov:1967pk} in his approach to gravity
as an `induced theory'. Indeed, focusing on the spin $2$ case
above, the effective action boils down to the computation
of the determinant of the conformal Laplacian
(since the scalar field action is quadratic). 
In practice, one needs to resort to a regularization scheme,
say for instance the use of a UV cut-off.
Expanding the effective action in powers of the cut-off,
several coefficients consist of local functionals
of the metric $g$ and its derivatives,
which are diffeomorphism-invariant.
The latter can be considered as potential actions 
for the metric, whose diffeomorphism invariance 
stems from a successful deformation of the linear
gauge symmetries generated by $\xi$ in \eqref{eq:linear}.
Weyl transformations, that is rescalings of the metric
and the scalar field $\phi$ for the form
\begin{equation}
    g \longmapsto \Omega^2\,g
    \qquad\text{and}\qquad
    \phi \longmapsto \Omega^{-\frac{n-2}2}\,\phi\,,
\end{equation}
for an arbitrary (but nowhere vanishing) parameter $\Omega$
also leave invariant the action \eqref{eq:action_conformal_sc}
and define an all order completion of the linear
gauge transformations generated by $\sigma$.
In even dimensions, only one term in the effective action
is also invariant under Weyl transformations,
namely the coefficient of the logarithmically divergent piece.
For $n=4$, this term is essentially the integral
of the Weyl tensor squared (up to a total derivative
and a topological term), which is the action for Weyl gravity.

The lessons of these low spin examples is that
one can leverage the existence of conserved currents
to derive an action for gauge fields introduced as sources
of the aforementioned currents. For a \emph{massless}
free scalar, the latter are also traceless which leads
to an additional, `Weyl-type', symmetry.
Insisting on preserving this linear gauge symmetry 
at the non-linear level, and after having integrated out
the scalar field, leads to a unique action
for the gauge fields under consideration.

This procedure generalizes to the higher-spin currents,
thereby producing a coupling of the original complex scalar
to a background of higher-spin gauge fields,
via a differential operator, covariant under the associated
higher-spin symmetries which define a non-linear completion
of the linear gauge transformations
\begin{equation}
    \delta_{\xi,\sigma} h_{\mu_1 \dots \mu_s}
    = \partial_{(\mu_1} \xi_{\mu_2 \dots \mu_s)}
    + \eta_{(\mu_1 \mu_2}\,\sigma_{\mu_3 \dots \mu_s)}\,,
\end{equation}
for all integers $s \geq 1$. These were identified
as the linear symmetries of \emph{conformal higher-spin
gravity} (CHSGra), a higher-spin generalization
of conformal (super)gravity proposed by Fradkin and Tseytlin 
\cite{Fradkin:1985am} at the free level, and studied further
at the cubic level \cite{Fradkin:1990ps}. Accordingly,
Tseytlin proposed to define conformal higher-spin gravity 
as the coefficient of the logarithmically divergent piece
of the effective action of a scalar field
in a higher-spin background \cite{Tseytlin:2002gz}.
However, as the spin $2$ case already illustrates,
working out \emph{perturbatively} the exact expression
of the relevant differential operator encoding this coupling 
for all spins $s > 2$ seems unrealistic.
This is not to say that with a perturbative approach
to this problem it is impossible to get/recover
manifestly (higher-spin) covariant objects. It allows one
to compute the conformal higher-spin gravity action
at the lowest orders, and confirm that the quadratic piece
is the expected one \cite{Fradkin:1985am}, as argued
in \cite{Tseytlin:2002gz} and worked out in details
in \cite{Bekaert:2010ky}
(see also \cite{Liu:1998bu} for a similar approach
with $\mathcal{N}=4$ super Yang--Mills theory,
and \cite{Bonezzi:2017mwr} from a worldline perspective).

A. Segal proposed an elegant solution to the problem of coupling
a (complex) scalar field to a background of higher-spin fields 
and computing its effective action, by resorting
to symbol calculus, and more generally, 
to deformation quantization \cite{Segal:2002gd}.
Without delving into technical details---that we shall review 
in the bulk of the paper---the idea is to translate
action and its gauge symmetries which formally read
\begin{equation}\label{eq:Segal}
    S[\phi, h_s]
    = \tfrac12\,\bra\phi\widehat{H}[h_s]\ket\phi\,,
    \qquad\qquad 
    \delta_\varepsilon \widehat{H}
    = \widehat{\varepsilon}^{\,\dagger} \circ \widehat{H}
    + \widehat{H} \circ \widehat{\varepsilon}\,,
    \qquad 
    \delta_\varepsilon \ket\phi
        = -\widehat{\varepsilon}\ket\phi\,,
\end{equation}
where $\widehat{H}$ and $\widehat{\varepsilon}$
are differential operators respectively encoding
the coupling to background fields $h_s$ and gauge parameters
(which appear as coefficients of these operators), 
into the language of symbols,
i.e. functions on the cotangent bundle $T^*\R^n$.
This approach has some computational advantages,
and in particular, the cubic part of the action
for CHSGra was derived \cite{Segal:2002gd} in this framework.

One of the drawback of both approaches outlined above,
however, is that they are defined around flat spacetime.
Working out the expression of conserved currents
for a free scalar field on a more general background
can be rather challenging, although the case of Weyl-flat
space (and $\mathcal{N}=1$ supersymmetrization thereof)
has been successfully worked out \cite{Kuzenko:2022hdv}.
More generally, formulating CHSGra around an arbitrary 
background or in a manifestly covariant manner, 
has been the subject of several works \cite{Grigoriev:2016bzl, Beccaria:2017nco, Kuzenko:2019ill, Kuzenko:2019eni, Kuzenko:2022hdv} (see also \cite{Kuzenko:2017ujh, Kuzenko:2021pqm, Kuzenko:2022qeq, Kuzenko:2024vms,Buchbinder:2024pjm} for supersymmetric extensions, and \cite{Joung:2021bhf}
for an approach to conformal gravity using `unfolding').

In this paper, we shall expand on the framework 
developed in \cite{Grigoriev:2016bzl} and \cite{Basile:2022nou},
wherein Segal's ideas were combined with techniques
from Fedosov quantization in order to obtain
a background-independent formulation of CHSGRa, 
and work out how to express the coupling of a scalar field
to a conformal higher-spin fields within this setting.
Having both the CHSGra action, and the action
for a matter scalar field coupled to it, in the same formalism
would allow one to probe various aspects of this coupling.

This paper is organized as follows: in Section \ref{sec:Fedosov}
we review the main constructions of Fedosov quantization
for the cotangent bundle of any manifold (our spacetime),
before introducing in Section \ref{sec:Wigner}
an analogue of the \emph{Wigner function} 
which we use to build an action for a scalar field
coupled to a background of higher-spin fields
in this framework. In Section \ref{sec:HS}, 
we illustrate our proposal by detailing the case
of the conformally-coupled scalar, and show
how Weyl symmetry is embedded in our formulation.
We also explain how higher-spin couplings arise,
together with the gauge symmetries. We conclude
the paper in Section \ref{sec:discu} with a discussion
of possible further directions to be explored,
and complement it with a short review of Weyl calculus
in Appendix \ref{app:Weyl_calculus}, some computational
details about Weyl transformations in the formalism
presented here in Appendix \ref{app:Weyl_transfo},
a quick review the definition of the FFS cocycle
in Appendix \ref{app:FFS}, and finally
a curvature expansion of the Fedosov connection
is presented in Appendix \ref{app:curvature}.

\section{Elements of Fedosov quantization}
\label{sec:Fedosov}
Before spelling out our action for a complex scalar
coupled to an arbitrary higher-spin background, 
we shall briefly review some constructions proposed
by Fedosov in his seminal paper \cite{Fedosov:1994zz}
on the deformation quantization of symplectic manifolds
(see also his textbook \cite{Fedosov:1996} for more details).
Readers familiar with these ideas may safely skip 
this section, while unfamiliar readers interested
in complementary references may consult
\cite[App. A]{Grigoriev:2016bzl} which we closely follow,
as well as \cite{Basile:2022nou} where these techniques
have been used in the context of conformal higher-spin gravity.

\paragraph{Building the Fedosov connection.}
The ingredient we need is a flat connection on the Weyl bundle,
\begin{equation}
    \WeylBundle_\Base := S(T\Base) \otimes \hat S(T^*\Base)
    \twoheadrightarrow \Base\,,
\end{equation}
where $\Base$ denotes our $n$-dimensional spacetime manifold,
and $\hat S(\dots)$ the completion of the symmetric algebra.
To be concrete, a typical section of this bundle
locally takes the form
\begin{equation}
    \Gamma(\WeylBundle_\Base)\ \ni\ {\tt a}(x;y,p)
    = \sum_{k,l} {\tt a}_{a_1 \dots a_k}^{b_1 \dots b_l}(x)\,
    y^{a_1} \dots y^{a_k}\,p_{b_1} \dots p_{b_l}\,,
\end{equation}
where $\{y^a\}$ and $\{p_b\}$, for $a,b=1,\dots,n:=\dim\Base$,
respectively define a basis of its cotangent
and tangent space over the point $x \in \Base$.
The above section is \emph{polynomial} in $p$, but is allowed
to be a \emph{formal power series} in $y$, in accordance 
with the fact that the Weyl bundle is the tensor product
of the symmetric algebra of $T\Base$, and the \emph{completion}
of the symmetric algebra of $T^*\Base$. 

The fiber at each point is isomorphic, upon extending it
over $\R\llbracket\hbar\rrbracket$, to that of the Weyl algebra 
$\WeylAlg_{2n}$ generated by the $2n$ variables $y$ and $p$,
whose associative (but non-commutative) product $\ast$
is given by
\begin{equation}
    \big(f \ast g\big)(y,p) = f(y,p)\,
    \exp\Big(\tfrac\hbar2\,
    \big[\tfrac{\overleftarrow{\partial}}{\partial y}
    \cdot \tfrac{\overrightarrow{\partial}}{\partial p}
    -\tfrac{\overleftarrow{\partial}}{\partial p}
    \cdot \tfrac{\overrightarrow{\partial}}{\partial y}\big]\Big)\,g(y,p)\,,
\end{equation}
where we denoted the contraction of Latin indices
by a dot, i.e. $y \cdot p = y^a\,p_a$.
This product is called the Moyal--Weyl product,
see Appendix \ref{app:Weyl_calculus} for a review
of its derivation from the perspective of symbol calculus.
Note that the operation\footnote{One can think
of it as essentially complex conjugation,
upon considering $\hbar$ as a \emph{purely imaginary}
formal parameter. When deriving the Moyal--Weyl product
from  the point of view of symbol calculus, as recalled
in Appendix \ref{app:Weyl_calculus}, the $\hbar$ factor
in its definition appears multiplied by the imaginary unit,
which we chose to absorb in $\hbar$ itself here
to simplify computations.}
\begin{equation}
    \hbar^\dagger = -\hbar\,,
    \qquad 
    (y^a)^\dagger = y^a\,,
    \qquad 
    (p_a)^\dagger = p_a\,,
\end{equation}
which also acts by complex conjugation on coefficients,
defines an anti-involution of the Weyl algebra, that is 
\begin{equation}
    (f \ast g)^\dagger = g^\dagger \ast f^\dagger\,,
\end{equation}
for any pair of elements $f$ and $g$.
The sections of the Weyl bundle can therefore be multiplied,
using the Moyal--Weyl product fiberwise, and thereby making 
$\WeylBundle_\Base$ into a bundle of associative algebras.
The Weyl algebra can be endowed with a grading, namely
\begin{equation}\label{eq:deg}
    \deg(y^a) = 1 = \deg(\hbar)\,,
    \qquad \deg(p_a) = 0\,,
\end{equation}
with respect to which the Moyal--Weyl product is of degree $0$.

Having recalled the definition of the Weyl bundle,
we can come back to our initial goal which is to construct
a flat connection on it. As it turns out, this is relatively
simple, as one can show that any $1$-form connection of the form
$A_0 = \dR x^\mu\,e_\mu^a\,p_a + \dots$,
where $e_\mu^a$ are the components of an invertible
frame field on $\Base$ and the dots denote higher order terms
in $y$ and $p$, can be extended into a flat connection
on $\WeylBundle_\Base$,
\begin{equation}
    \dR A + \tfrac1{2\hbar}\,[A,A]_\ast = 0\,,
    \qquad\text{with}\qquad 
    A = A_0 + \text{(corrections)}\,.
\end{equation}
A simple way of constructing such a flat connection
is to start from
\begin{equation}
    A_0 = \dR x^\mu\,\big(e_\mu^a\,p_a
    + \omega_\mu^{a,b}\,p_a\,y_b\big)\,,
\end{equation}
where $\omega^{a,b} := \dR x^\mu\,\omega_\mu^{a,b}$
are the components of the torsionless spin-connection
with respect to the vielbein $e^a_\mu$, which preserves
the fiber metric $\eta^{ab}$, used to raise and lower
the fiber (i.e. Latin) indices. Let us introduce
\begin{equation}
    \delta := -\tfrac1\hbar\,[\dR x^\mu\,e^a_\mu\,p_a,-]_\ast\,,
    \qquad 
    \nabla := \dR + \tfrac1\hbar\,
    [\dR x^\mu\,\omega_\mu^{a,b}\,p_a\,y_b, -]_\ast\,,
    \qquad
    R^\nabla := \big(\dR\omega^{a,b}
    + \omega^{a,}{}_c\,\omega^{c,b}\big)\,p_a\,y_b\,,
\end{equation}
so that the curvature of $\nabla$ is simply given by
\begin{equation}
    \nabla^2 = \tfrac1\hbar\,[R^\nabla,-]_\ast\,,
\end{equation}
and one can easily check that
\begin{equation}
    \delta\nabla + \nabla\delta = 0\,,
\end{equation}
as a consequence of the torsionlessness of $\nabla$.
Note that $\delta$ and $\nabla$ are respectively
of degree $-1$ and $0$ with respect to the previously 
introduced grading \eqref{eq:deg}.
One can then show that there exists a unique $1$-form
$\Completion \in \Omega^1(\Base, \WeylBundle_\Base)$
of degree $\geq 2$ such that
\begin{equation}\label{eq:Fedosov}
    A = A_0 + \Completion\,,
\end{equation}
defines a \emph{flat} connection on the Weyl bundle,
with $\Completion$ linear in $p$ and obeying $h\Completion=0$,
and where
\begin{equation}\label{eq:defh&N}
    h := \tfrac1N\,y^a\,e_a^\mu\,\tfrac{\partial}{\partial (\dR x^\mu)}\,,
    \qquad\qquad
    N := y^a\,\tfrac{\partial}{\partial y^a}
        + \dR x^\mu\,\tfrac{\partial}{\partial (\dR x^\mu)}\,,
\end{equation}
with $N$ the number operator returning the sum
of the form degree and $y$-degree of its argument.
Equivalently, the associated covariant derivative
\begin{equation}
    \Fedosov := \dR + \tfrac1\hbar\,[A,-]_\ast
    \equiv -\delta + \nabla + \tfrac1\hbar[\Completion,-]_\ast\,,
\end{equation}
defines a \emph{differential}, i.e. squares to zero,
on the Weyl bundle. The $1$-form $\Completion$
can be computed order by order in $y$ via the recursive formulae
\begin{equation}\label{eq:recA}
    \Completion_{(2)} = h(R^\nabla)
    \qquad\text{and}\qquad
    \Completion_{(k+1)} = h\Big(\nabla \Completion_{(k)}
    + \tfrac1{2\hbar}\,\sum_{l=2}^{k-1}
    [\Completion_{(l)}, \Completion_{(k+1-l)}]_\ast\Big)
    \quad\text{for}\quad k\geq2\,,
\end{equation}
which yield
\begin{equation}
    \begin{aligned}
        \Completion
        & = - \tfrac13\,\dR x^\mu\,R_{\mu a}{}^c{}_b\,y^a y^b p_c
        - \tfrac1{12}\,\dR x^\mu\,\nabla_a R_{\mu b}{}^d{}_c\,y^a y^b y^c p_d \\
        & \hspace{50pt}
        -\dR x^\mu\,\big[\tfrac1{60}\,\nabla_a \nabla_b R_{\mu c}{}^e{}_d
        + \tfrac1{45}\,R_{\times a}{}^e{}_b\,
        R_{\mu c}{}^\times{}_d\big]\,y^a y^b y^c y^d p_e + (\cdots)
    \end{aligned}
\end{equation}
where the dots denote terms of higher order in $y$.%
\footnote{Remark that the grading \eqref{eq:deg}
with respect to which the defining recursion relation
for $\Completion$ is given, reduces to the degree
of homogeneity in $y$. This is a consequence of the fact
that the first correction $\Completion_{(2)}$ is linear in $p$
so that, not only all higher order correction stay linear in $p$, 
but also the star-commutator in \eqref{eq:recA} reduces
to the Poisson bracket piece, i.e. to its piece
of order $\hbar^1$. Consequently, no $\hbar$ correction
appear in $\Completion$ in the case of interest here.}
Introducing the notation
\begin{equation}
    \Curvature := hR^\nabla\,,
    \qquad\text{and}\qquad 
    \SymDer := h\nabla\,,
\end{equation}
we can re-sum the defining relations of $\Completion$ as
\begin{equation}\label{eq:rec_A}
    \Completion = \Curvature + \SymDer \Completion
    + \tfrac1{2\hbar}\,h[\Completion,\Completion]_\ast\,,
\end{equation}
so that the first few orders of $\Completion$ in $y$
can be re-written as
\begin{equation}
    \Completion_{(2)} = \Curvature\,,
    \quad \Completion_{(3)} = \SymDer\Curvature\,,
    \quad \Completion_{(4)} = \SymDer^2\Curvature
    + \tfrac1{2\hbar}\,h\big[\Curvature,\Curvature\big]_\ast\,.
\end{equation}

As mentioned above, any $1$-form connection
valued in the Weyl algebra whose component along $p_a$
is an invertible vielbein can be extended to a flat connection
by the same mechanism as above: the vielbein piece
gives rise to the differential $\delta$, and the components
of the $1$-form valued in the Weyl bundle needed to flatten
the original connection can be computed recursively
using its contracting homotopy $h$. In particular,
one may start from a connection containing higher-spin
components which appear as terms of higher order in $p$
(and $y$) in the initial data, e.g.
\begin{equation}\label{hsconnec}
    A_0 = e^a\,p_a + \omega^{a,b}\,p_a\,y_b
    + e^{ab}\,p_a p_b + \omega^{ab,c}\,p_a p_b\,y_c + \dots\,,
\end{equation}
and find $\Completion$ so that $A=A_0+\Completion$
is flat, though the $1$-form $\Completion$ will also involve
the curvature of these higher-spin components.%
\footnote{Note that in the case where the initial data
$A_0$ contain higher-spin components (higher orders in $p$),
one should use a slightly different degree, namely
one should assign degree $1$ to both $y$ and $p$
and degree $2$ to $\hbar$, so that the Moyal--Weyl product
remains of degree $0$ with respect to this new grading.
This is actually the gradation used originally by Fedosov
\cite{Fedosov:1994zz, Fedosov:1996}, for more details
see also, e.g., \cite[App. E]{Basile:2022nou}.} The higher components of \eqref{hsconnec} correspond to vielbeins and spin-connections of conformal higher-spin fields within the frame-like formulation, which was developed in \cite{Fradkin:1989md,Vasiliev:2009ck}.

\paragraph{Lift of symbols and invariant trace.}
Once the Fedosov connection $\Fedosov$ is constructed,
we can define the lift of the symbol
of a differential operator on $\Base$, that is a function
on the cotangent bundle $T^*\Base$, say
$f(x,p) \in \Functions_{pol.}(T^*\Base) \cong \Gamma(ST\Base)$,
as the (unique) section
$F(x;y,p) \in \Gamma(\Base,\WeylBundle_\Base)$ verifying
\begin{equation}
    \Fedosov F = 0\,, 
    \qquad 
    F\rvert_{y=0} = f\,,
\end{equation}
i.e. the (unique) covariantly constant section
of the Weyl bundle whose order $0$ in $y$ is $f$.
In other words, starting from a function
only of $x^\mu$ and $p_a$, one reconstruct
a flat section of the Weyl bundle, which is a function
of $x^\mu$, $p_a$ and $y^a$, whose dependency on $y$
is completely determined by the covariant constancy condition,
and the coefficients of these terms proportional to $y$
are obtained from the original function of $x$ and $p$.
To do so, one simply needs to solve the covariant
constancy condition, which can be done iteratively via
\begin{equation}
    \label{eq:rec_F}
    F_{(0)} = f
    \qquad\text{and}\qquad
    F_{(k+1)} = h\Big(\nabla F_{(k)} + \tfrac1\hbar\,
    \sum_{l=2}^{k+1} [\Completion_{(l)}, F_{(k+1-l)}]_\ast\Big)
    \quad\text{for}\quad k \geq 0\,,
\end{equation}
where $F_{(n)}$ denotes the component of the lift $F$ 
homogeneous of degree $n$ with respect to grading \eqref{eq:deg},
i.e. it corresponds to the homogeneity both in $y$ and $\hbar$.
This leads to
\begin{equation}\label{eq:lift_y2}
    F(x;y,p) = f + y^a\,\nabla_a f + \tfrac12\,y^a y^b\,
    \big(\nabla_a \nabla_b + \tfrac13\,R_{da}{}^c{}_b\,
    p_c\,\tfrac{\partial}{\partial p_d}\big)\,f
    + (\dots)\,,
\end{equation}
at the first few orders. This lift of (fiberwise polynomial) 
functions establishes a bijection between the latter
and covariantly constant sections of the Weyl bundle, 
\begin{equation}
    \begin{aligned}
        \tau: \Functions(T^*\Base)\
        & \overset{\sim}{\longrightarrow}\
        {\rm Ker}(\Fedosov) \subset \Gamma(\WeylBundle_\Base) \\
        f(x,p)\,\ & \longmapsto\ F(x;y,p)
                                \equiv \tau(f)(x;y,p)\,,
    \end{aligned}
\end{equation}
and allows us to define a \emph{star-product},
i.e. an associative but non-commutative deformation
of the pointwise product, via the simple formula%
\footnote{Remark that, by construction, the evaluation
of a covariantly constant section at $y=0$ yields
the function on the cotangent bundle that it is the lift of.
In other words, this simple operation is the inverse
of the lift $\tau$, i.e. $\tau^{-1}(-) = (-)\rvert_{y=0}$.
The star-product on $T^*\Base$ can therefore be written as
$f \star g = \tau^{-1}\big(\tau(f) \ast \tau(f)\big)$
which makes it clear that the lift $\tau$ is a morphism
of algebras between $\big({\rm Ker}(\Fedosov), \ast\big)$
and $\big(\Functions(T^*\Base), \star\big)$,
the star-product on the latter being `pulled-back'
from the Moyal--Weyl one defined fiberwise.}
\begin{equation}\label{eq:star-product}
    f \star g = (F \ast G)\big|_{y=0}\,,
    \qquad 
    f, g \in \Functions_{pol}(T^*\Base)\,,
\end{equation}
where $F, G \in \Gamma(\WeylBundle_\Base)$ are the lifts
of $f$ and $g$ respectively. Associativity simply follows
from the fact that the Moyal--Weyl product in the fiber
is itself associative. To summarize, we are able to define
a star-product on the cotangent bundle of our spacetime 
$T^*\Base$ thanks to the fact that any function can be lifted 
to a flat section of the Weyl bundle, wherein we can use
the Moyal--Weyl star-product to multiply the flat sections 
corresponding to two functions on $\Base$, and evaluate
the result at $y=0$ thereby producing another function
on $T^*\Base$. 

There exists a trace (essentially unique) on the space
of covariantly constant sections of the Weyl bundle,
which takes the form \cite{Basile:2022nou}
\begin{equation}
    \Tr_A(F) = \int_{x \in \Base} \int_{T^*_x\Base} \dR^np\ 
    \mu(F|\underbrace{A,\cdots,A}_{n\,\text{times}})\,,
\end{equation}
where $\mu: \WeylAlg_{2n}^{\wedge n} \otimes \WeylAlg_{2n}
\longrightarrow \R[p_a]$ is a multilinear map
valued in polynomials in $p$,
obtained from the Feigin--Felder--Shoikhet cocycle
\cite{Feigin:2005}. The fact that $\mu$ is obtained
from a Hochschild cocycle for the Weyl algebra ensures
two important properties of this trace: it is invariant
under the gauge transformations
\begin{equation}
    \delta_\xi A = \dR\xi + \tfrac1\hbar\,[A,\xi]_\ast\,,
    \qquad 
    \xi \in \Gamma(\WeylBundle_\Base)\,,
\end{equation}
of the flat connection $A$ up to boundary terms, i.e.
\begin{equation}
    \delta_\xi \Tr_A(F) = \int_\Base \int \dR^np\
    \big[\dR (\dots) + \tfrac{\partial}{\partial p_a} (\dots)_a\big]\,,
\end{equation}
and it is cyclic, also up to boundary terms, 
\begin{equation}
    \Tr_A([F, G]_\ast) = \int_\Base \int \dR^np\
    \big[\dR (\dots) + \tfrac{\partial}{\partial p_a} (\dots)_a\big]\,,
\end{equation}
for any covariantly constant sections $F$ and $G$.

The detailed expression for $\mu$
is given in Appendix \ref{app:FFS}, for the moment
it is enough for our purpose to know that,
for a flat connection $A$ which is linear in $p$
as the example reviewed previously,
the associated trace of any lifted symbol $F$
boils down to
\begin{equation}\label{eq:trace_linear_case}
    \Tr_A(F) = \int_\Base \dR^nx\,|e|\,
    \int_{T^*_x\Base} \dR^np\ \sum_{k\geq0} 
    \mu^{\scriptscriptstyle\nabla}_{a_1 \dots a_k}(x)\,
    \frac{\partial^k}{\partial p_{a_1} \dots \partial p_{a_k}}
    F\big|_{y=0}\,,
\end{equation}
where $\mu^{\scriptscriptstyle\nabla}_{a_1 \dots a_k}(x)$
are polynomials in the curvature of $\nabla$
and covariant derivatives thereof. 

What has been reviewed above is just the Fedosov approach
to deformation quantization for the particular case
of the symplectic manifold being the cotangent bundle
(of the spacetime), which was also studied by Fedosov himself 
\cite{Fedosov2001}. A development since \cite{Fedosov2001}
is the construction of the invariant trace
by Feigin, Felder and Shoikhet \cite{Feigin:2005}.
Let us briefly explain now, see \cite{Basile:2022nou}
for more details, how this is related to conformal
higher-spin fields. To begin with, an off-shell description
of conformal higher-spin fields requires
the Fedosov connection $A$ and a covariantly constant section
of the Weyl bundle $F$. Different types of scalar matter,
i.e. whether we start out with $\mathcal{L}\sim\phi \square \phi$ 
or $\mathcal{L}\sim\phi \square^k \phi$, $k>1$,
lead to different spectra of (higher-spin) currents and,
hence, to different spectra of sources/background
(higher-spin) fields. An immediate consequence
is that it is necessary to fix the background value for $F$
to land on a specific theory. We will consider $F$ of the form
\begin{align}
    F & = p^2 + \sum_{s>2} h^{a_1 \dots a_s}(x)\,
    p_{a_1} \dots p_{a_s} + \dots\,,
\end{align}
where the presence of $p^2$ here implies that we are coupling
the usual free scalar field $\mathcal{L}\sim\phi \square \phi$
to (higher-spin) background fields $h^{a_1 \dots a_s}$.
The formalism is flexible enough to allow one
to realize conformal higher-spin fields both in the frame-like, 
cf. \eqref{hsconnec}, and in the metric-like ways, as below.
In this paper, we prefer to keep $A$ purely gravitational,
i.e. it is completely expressed in terms of a vielbein $e^a$. 
With the help of the $\xi$ gauge-symmetry one can move
between the frame-like and metric-like formulations.

\paragraph{Fock space bundle.}
Having constructed a bundle of Weyl algebra, let us now 
proceed with the definition of a vector bundle associated
with the Fock representation. As a vector space, the latter
can be identified with the subspace of 
$\WeylAlg_{2n} \cong \R[y^a, p_b]$ consisting of 
polynomials (or even formal power series) in $y$,
that we shall denote by $\Fock_n \equiv \R[y^a]$.
The representation is given by the \emph{quantization map},
\begin{equation}
    \big(\rho(f) \varphi\big)(y)
    = f(y,p)\,\exp\Big(\!-\hbar\,
    \tfrac{\overleftarrow{\partial}}{\partial p} \cdot 
    \big[\tfrac12\,\tfrac{\overleftarrow{\partial}}{\partial y}
    +\tfrac{\overrightarrow{\partial}}{\partial y}\big]\Big)\,
    \varphi(y)\big|_{p=0}\,,
\end{equation}
for any element $f(y,p) \in \WeylAlg_{2n}$ of the Weyl algebra 
and $\varphi(y) \in \Fock_n$ of the Fock space. That it defines
a representation of the Weyl algebra means that it verifies
\begin{equation}
    \rho(f) \circ \rho(g) = \rho(f \ast g)\,,
    \qquad 
    f, g \in \WeylAlg_{2n}\,.
\end{equation}
The name `quantization map' comes from the fact
that it allows one to associate, to any (polynomial)
function of $\R^{2n} \cong T^*\R^n$, which are nothing but
elements of the Weyl algebra, a differential operator 
acting on the space of `wave functions',
i.e. smooth functions on $\R^n$, which we consider
as elements of the Fock space (via for instance
their Taylor series). Put differently, the pair
$(\WeylBundle_{2n}, \Fock_n)$ can be thought of
as a \emph{flat model} for the quantization
of a cotangent bundle $T^*\Base$ with $\dim\Base=n$,
wherein the Weyl algebra models the algebra of functions
on $T^*\Base$, while the Fock space models smooth functions
on the base manifold $\Base$, on which functions
on the cotangent bundle act as differential operators.

Given now an arbitrary smooth manifold $\Base$,
one can consider the `bundle of Fock spaces'
defined as
\begin{equation}
    \Fock_\Base := S(T^*\Base)
                    \twoheadrightarrow \Base\,,
\end{equation}
whose sections are
\begin{equation}
    \Gamma(\Fock_\Base)\, \ni\, \Phi(x;y)
    = \sum_{k\geq0} \tfrac1{n!}\,\Phi_{a_1 \dots a_k}(x)\,
    y^{a_1} \dots y^{a_k}\,,
\end{equation}
that we shall extend as formal power series in $\hbar$.
A Fedosov connection $A$ defines a flat covariant derivative
on this Fock bundle, whose local expression is
\begin{equation}
    \Fedosov = \dR + \tfrac1\hbar\,\rho(A)\,,
\end{equation}
with $\rho$ is the quantization map above.
A simple computation leads to
\begin{equation}
    \rho(p_a) = -\hbar\,\tfrac{\partial}{\partial y^a}\,,
    \qquad 
    \rho(y^a p_b) = -\tfrac\hbar2\,
    \big(y^a\,\tfrac{\partial}{\partial y^b} + \tfrac{\partial}{\partial y^b}\,y^a\big)
    = -\hbar\,\big(y^a\tfrac{\partial}{\partial y^b}
        + \tfrac12\,\delta^a_b\big)\,,
\end{equation}
and more generally, 
\begin{equation}
    -\tfrac1\hbar\,\rho(y^{a_1} \dots y^{a_n} p_b)
    = y^{a_1} \dots y^{a_n}\,\tfrac{\partial}{\partial y^b}
    + \tfrac{n}2\,\delta_b^{(a_1} y^{a_2} \dots y^{a_n)}\,,
\end{equation}
so that, upon choosing $\nabla$ to be a metric connection,
and $A$ the flat connection \eqref{eq:Fedosov} built from it
as explained above, one finds
\begin{equation}
    \Fedosov\Phi = \Big(\!-\delta + \nabla
                    + \rho(\Completion)\Big)\Phi\,,
    \qquad\text{with}\qquad 
    \delta = e^a\,\tfrac{\partial}{\partial y^a}\,,
\end{equation}
which in particular, contains the \emph{same acyclic piece}
$\delta$ as in the Fedosov connection \eqref{eq:Fedosov},
which is an operator of degree $n-1$ in $y$. As a consequence,
we can solve for covariantly constant sections
of the Fock bundle in a similar manner as we did 
in the Weyl bundle: expanding the condition
\begin{equation}
    \Fedosov\Phi = 0\,,
    \qquad\text{with}\qquad 
    \Phi\big|_{y=0} = \phi\,,
\end{equation}
order by order in $y$ yields
\begin{equation}
    \delta\Phi_{(n+1)} = \nabla\Phi_{(n)}
    + \tfrac1\hbar\,\sum_{k=2}^{n+1} \rho(\Completion_{(k)})
    \Phi_{(n+1-k)}\,,
\end{equation}
which gives us a definition of the order $n+1$ term
in the $y$ expansion of $\Phi$ thanks to the contracting
homotopy \eqref{eq:defh&N} introduced before, i.e.
\begin{equation}\label{eq:rec_phi}
    \Phi_{(n+1)} = h\Big(\nabla\Phi_{(n)}
    + \tfrac1\hbar\,\sum_{k=2}^{n+1} \rho(\Completion_{(k)})
    \Phi_{(n+1-k)}\Big)\,.
\end{equation}
The whole covariant section $\ket\Phi$ only depends
on its value at $y=0$, which is a function on $\Base$,
thereby establishing a bijection
\begin{equation}
    \begin{aligned}
        \tau: \Functions(\Base)\
        & \overset{\sim}{\longrightarrow}\
        {\rm Ker}(\Fedosov) \subset \Gamma(\Fock_\Base) \\
        \phi(x)\,\ & \longmapsto\ \Phi(x;y)
                                \equiv \tau(\phi)(x;y)\,,
    \end{aligned}
\end{equation}
between $\Functions(\Base)$ and covariantly constant sections 
of the Fock bundle (that we denoted by the same symbol $\tau$
as the isomorphism between functions on the cotangent bundle
and flat sections of Weyl bundle,
in a slight abuse of notation).
The first few order of the covariantly constant section
associated with $\phi(x)$ read
\begin{equation}\label{eq:lift_phi}
    \Phi(x;y) = \phi + y^a\nabla_a\phi
    + \tfrac12\,y^a y^b\,\big(\nabla_a\nabla_b
    - \tfrac16\,R_{ab}\big)\phi + \dots
\end{equation}
where $R_{ab}$ denotes the Ricci tensor of $\nabla$,
and the dots denote terms of order $3$ or higher in $y$.

We can now define a quantization map in this curved setting,
that is to say, a way to associate to any symbol
$f \in \Functions_{pol}(T^*\Base)$, that is any fiberwise 
polynomial function on the cotangent bundle of $\Base$,
a differential operator $\widehat f$
which acts on `wave functions', i.e. functions
$\phi\in\Functions(\Base)$ on the base, defined as follows%
\footnote{Let us note that, as for the star-product,
writing the quantization map in terms of the lift $\tau$,
namely $\widehat{f}\phi
= \tau^{-1}\big[\rho\big(\tau(f)\big)\tau(\phi)\big]$,
makes it apparent that the latter defines a morphism
of pairs algebra-module between
$\big(\Functions(T^*\Base), \Functions(\Base)\big)$
and flat sections of the Weyl and Fock bundles.
This also shows that the quantization map
on $\Functions(\Base)$ is `pulled-back' from that 
on flat sections of the Fock bundle, in complete parallel
with the definition of the star-product on $T^*\Base$.}
\begin{equation}
    \big(\widehat f\phi\big)(x)
        := \rho(F) \Phi \big|_{y=0}\,,
\end{equation}
where $F \in \Gamma(\WeylBundle_\Base)$
and $\Phi \in \Gamma(\Fock_\Base)$ are the lifts
of $f$ and $\phi$ respectively as a covariantly constant
section of the Weyl and Fock bundles. This defines
a representation of the star-product algebra
$\big(\Functions_{pol}(T^*\Base), \star\big)$
on the space of `wave functions' $\Functions(\Base)$, i.e.
\begin{equation}
    \widehat f \circ \widehat g = \widehat{f \star g}
    \qquad
    \forall f, g \in \Functions_{pol}(T^*\Base)\,,
\end{equation}
where $\star$ is the star-product defined
in \eqref{eq:star-product}. Here again, this is
simply a consequence of the fact that $(\Fock_n,\rho)$
is a representation of $(\WeylAlg_{2n},\ast)$,
i.e. we sort of `pullback' the algebra and representation
structure from the fiber to the base manifold $\Base$.
Note that this approach was already outlined in
\cite[App. A]{Grigoriev:2016bzl}.

\section{Wigner function and quadratic actions}
\label{sec:Wigner}
We have now all the ingredients needed to re-express
the coupling of a scalar field to an arbitrary higher-spin
background. The latter is encoded in a pair of fields,%
\footnote{Such a description is obtained from an approach
known as the `parent formulation' of gauge theories,
developed in \cite{Barnich:2004cr, Barnich:2010sw, Grigoriev:2010ic, Grigoriev:2012xg} and references therein.}
namely a flat connection $A$ and a covariantly constant 
section $F$ of the Weyl bundle \cite{Grigoriev:2006tt, Grigoriev:2016bzl},
\begin{equation}
    \dR A + \tfrac1{2\hbar}\,[A,A]_\ast = 0\,,
    \qquad
    \dR F + \tfrac1\hbar\,[A,F]_\ast = 0\,,
\end{equation}
which is invariant under the gauge transformations
\begin{equation}
    \delta_\xi A = \dR\xi + \tfrac1\hbar\,[A,\xi]_\ast\,,
    \qquad 
    \delta_{\xi,w} F = \tfrac1\hbar\,[F,\xi]_\ast
    + \{F,w\}_\ast\,,
    \qquad 
    \dR w + \tfrac1\hbar\,[A,w]_\ast \overset{!}{=} 0\,,
\end{equation}
where $\xi,w \in \Gamma(\WeylBundle_\Base)$
are $0$-form valued in the Weyl bundle,
with $w$ required to be covariantly constant,
while $\xi$ is unconstrained. The sum
$\varepsilon=\tfrac1\hbar\xi+w$ corresponds
to the symbol of an arbitrary differential operator,
such as $\widehat{\varepsilon}$ appearing in \eqref{eq:Segal},
and it splits into its Hermitian and anti-Hermitian part,
respectively $w$ and $\xi$ (though both are real, 
the latter is dressed with $\hbar$ that we take as imaginary
in the sense that $\hbar^\dagger=-\hbar$).

As usual when dealing with gauge theories, matter fields
consists of sections of vector bundles associated with
representation of the gauge algebra (meaning here,
the algebra in which gauge fields take values).
Accordingly, we add the scalar field $\phi$ 
to the previous system in the guise of its lift
as a covariantly constant section of the Fock bundle,
\begin{equation}
    \dR\Phi + \tfrac1\hbar\,\rho(A)\Phi = 0\,,
\end{equation}
which transforms in the corresponding representation, 
\begin{equation}
    \delta_{\xi,w} \Phi = -\rho(\tfrac1\hbar\,\xi + w)\Phi\,,
\end{equation}
thereby preserving the covariant constancy condition.
Now all we need is an action functional implementing
the coupling of $\phi$ to the higher-spin background
in a gauge-invariant manner.

Around flat space, Segal's approach consisted in
considering a quadratic action for a \emph{complex}
scalar field $\phi$ in flat spacetime, 
\begin{equation}
    S[\phi] = \int_{\R^n} \dR^nx\ \phi^*(x)\,
                            (\widehat H\phi)(x)\,,
\end{equation}
for some differential operator $\widehat{H}$
which encode the coupling of $\phi$ to a background 
of gauge fields, the latter being related
to the `coefficients' of this operator. For instance,
in the case of the conformally-coupled scalar,
$\widehat{H}$ would be the conformal Laplacian
whose expression depends on a metric $g$ (via its inverse
contracting two covariant derivatives,
and via the Ricci scalar term), and which implements
the coupling of $\phi$ to conformal gravity.
The above action can formally be written as
\begin{equation}
    S[\phi] = \bra{\phi} \widehat{H} \ket{\phi}\,,
\end{equation}
so that it becomes relatively simple to see that
it is invariant under the following
infinitesimal transformations
\begin{equation}
    \delta_{\varepsilon}\ket{\phi}
    = -\widehat \varepsilon\ket{\phi}\,,
    \qquad
    \delta_{\varepsilon} \widehat H
    = \widehat{\varepsilon}^{\,\dagger} \circ \widehat{H}
    + \widehat{H} \circ \widehat{\varepsilon}\,,
\end{equation}
where $\varepsilon$ is another, arbitrary,
differential operator. Assuming that the space of operators
we are working with possesses a \emph{trace},
we can further re-write the action as
\begin{equation}
    S[\phi] = \Tr(\widehat{H} \circ \ket{\phi}\!\bra{\phi})\,,
\end{equation}
that is the trace of the operator $\widehat H$ composed
with the projector $\ket{\phi}\!\bra{\phi}$. In this form,
the action can be more easily translated in terms
of symbols, leading to
\begin{equation}
    S[\phi] = \int_{T^*\R^n} \dR^n p\,\dR^n x\
    \big(H \star W_{\phi}\big)(x,p)
\end{equation}
where $H(x,p)$ and $W_{\phi}(x,p)$ are the symbols
of the kinetic operator $\widehat{H}$ and the projector
$\ket{\phi}\!\bra{\phi}$, also known as the Wigner function,
respectively. The integration over the cotangent bundle
$T^*\R^n$ defines a trace over the space of symbols,
at least those which are compactly supported
or vanish at infinity sufficiently fast.
Indeed, in this case one finds
\begin{equation}
    \Tr(f \star g) = \int_{T^*\R^n} \dR^n x\,\dR^n p\
    (f \star g)(x,p) = \int_{T^*\R^n} \dR^n x\,\dR^n p\
    f(x,p)\,g(x,p) = \Tr(g \star f)\,,
\end{equation}
for any symbols $f$ and $g$, since all higher order terms
in the star product are total derivatives on $T^*\R^n$,
and hence can be ignored for the aforementioned
suitable class of symbols. The transformation rule,
in terms of symbols, becomes 
\begin{equation}
     \delta_{\varepsilon} H
     = \varepsilon^{\dagger} \star H + H \star \varepsilon\,,
    \qquad\text{and}\qquad
    \delta_{\varepsilon} W_{\phi} = -\varepsilon \star W_{\phi}
    - W_{\phi} \star \varepsilon^{\dagger}\,,
\end{equation}
under which the action transform as 
\begin{align}
    \delta_{\varepsilon} S[\phi]
    = -\Tr\big([H \star W_{\phi}\,,\varepsilon^{\dag}]_{\star}\big)
    = 0\,,
\end{align}
i.e. the action is left invariant as a consequence
of the cyclicity of the trace.

We have seen in the previous section how to define
a star-product and construct the associated invariant
trace via the FFS cocycle for any, possibly curved,
manifold $\Base$ so that we only need to find a suitable
generalization of the Wigner function to curved settings.
One can think of the Wigner function as a bilinear map
\begin{equation}
    W: \Fock_n \otimes \Fock_n \longrightarrow \WeylAlg_{2n}\,,
\end{equation}
taking two elements of the Fock representation
and constructing an element of the Weyl algebra
out of them. For our purpose, what matters is that
it possesses the following couple of properties
(whose proof are recalled in Appendix \ref{app:Weyl_calculus}).
\begin{enumerate}[label=$(\roman*)$]
\item\label{item:left-right_Wigner}
First, it intertwines the left and right multiplication
in the Weyl algebra with the Fock action
\begin{equation}
    F \ast W[\Phi,\Psi] = W[\rho(F)\Phi, \Psi]\,,
    \qquad 
    W[\Phi, \Psi] \ast F^\dagger = W[\Phi, \rho(F)\Psi]\,,
\end{equation}
for any element $F(y,p) \in \WeylAlg_{2n}$
and any pair of Fock space states
$\Phi(y), \Psi(y) \in \Fock_n$.

\item\label{item:integral_Wigner}
Second, integrating it over momenta yields
\begin{equation}
    \int_{\R^n} \dR^np\
    \tfrac{\partial^k}{\partial p_{a_1} \dots \partial p_{a_k}}\,
    W[\Phi,\Psi] = \delta_{k,0}\,\Phi(y)\,\Psi(y)\,,
\end{equation}
for any Fock space elements $\Phi,\Psi \in \Fock_n$ 
which are seen as embedded in the Weyl algebra
on the right hand side. 
\end{enumerate}

A first naive guess for a curved version $\Wigner_\phi$
of the Wigner function associated with a scalar field
$\phi \in\Functions(\Base)$ is to simply apply
the above bilinear map to two copies of its lift
as covariantly constant sections of the Fock bundle,
i.e.
\begin{equation}
    \Wigner_\phi(x;y,p) := W[\Phi,\Phi]
    = \int \dR^nu\ e^{\frac1\hbar\,p \cdot u}\,
    \Phi(x;y+\tfrac12\,u)\,\Phi^\dagger(x;y-\tfrac12\,u)\,.
\end{equation}
First of all, let us note that this is a covariantly
constant section of the Weyl bundle. Indeed, upon writing
it as $\Wigner_\phi = W[\Phi,\Phi]$ in order to highlight
the fact that it is bilinear in the covariantly constant
section of the Fock bundle $\Phi$, one finds that it verifies
\begin{equation}
    \tfrac1\hbar\,[A, \Wigner_\phi]_\ast
    = \tfrac1\hbar A \ast W[\Phi,\Phi]
    + W[\Phi,\Phi] \ast (\tfrac1\hbar\,A)^\dagger
    = W[\rho\big(\tfrac1\hbar\,A)\Phi,\Phi]
    + W[\Phi,\rho(\tfrac1\hbar\,A)\Phi]\,,
\end{equation}
where we used the properties \ref{item:left-right_Wigner}.
We can then use the covariant constancy of $\Phi$,
to show that
\begin{equation}
    \dR\Wigner_\phi + \tfrac1\hbar\,[A,\Wigner_\phi]_\ast = 0\,,
\end{equation}
i.e. our curved version the Wigner function $\Wigner_\phi$
is a covariantly constant section of the Weyl bundle.
Moreover, properties \ref{item:left-right_Wigner}
also ensure that $\Wigner_\phi$ transforms as
\begin{equation}
    \delta_{\xi,w} \Wigner_\phi
    = \tfrac1\hbar\,[\Wigner_\phi, \xi]_\ast
    - \{\Wigner_\phi,w\}_\ast\,.
\end{equation}
which implies that its star-product
with the covariantly constant lift $F$ behaves as
\begin{equation}
    \delta_{\xi,w}(F \ast \Wigner_\phi)
    = [F \ast \Wigner_\phi, \tfrac1\hbar\,\xi - w]_\ast\,,
\end{equation}
under the gauge transformations of the system.
As a consequence, the functional%
\footnote{Note that the dependence on conformal higher spin fields
in the action \eqref{main} is a little subtle to read-off:
as explained in \cite{Basile:2022nou}, they can be moved around
between $A$ and $F$ via gauge transformations,
which can therefore encode field redefinitions.}
\begin{equation}\label{main}
    S[\phi] = \Tr_A(F \ast \Wigner_\phi)\,,
\end{equation}
is well-defined, being the trace of the star-product
of two covariantly constant sections of the Weyl bundle,
as well as gauge invariant under all transformations
listed above thanks to the cyclicity of the FFS trace,
which holds \emph{up to boundary terms}. Let us remark that,
contrary to the action for CHS gravity which is expressed
as the FFS trace of a symbol that dies off at infinity
\emph{both} in spacetime and in the fiber/momenta directions 
\cite{Basile:2022nou}, this is not necessarily the case here:
the $p$-dependency of the integrand may not allows us to discard
boundary terms for arbitrary gauge parameters. In other words,
we expect that the gauge parameters $\xi$ and $w$ should be
restricted so as to ensure that the boundary terms appearing
when checking the cyclicity/gauge invariance of the FFS trace
(see \cite[App. C]{Basile:2022nou}) can actually be neglected.
Modulo this subtlety, eq. \eqref{main} gives a manifestly covariant 
and higher-spin invariant form of a coupling
between the scalar field and a background
of conformal higher-spin fields, which is one of the main results
of the paper. On the other hand, irrespectively of the action principle, the equations of motion $\rho(F) \Phi \big|_{y=0}=0$ are well-defined and, in particular, gauge invariant. As it turns out, in the case where $A$ is linear
in $p$, this expression simplifies to
\begin{equation}
    S[\phi] = \int_\Base \dR^nx\,|e|\,\int_{T^*_x\Base} \dR^np\
    W[\rho(F)\Phi,\Phi]\big|_{y=0}
    = \int_\Base \dR^nx\,|e|\,\phi^*(x)\,(\widehat f\phi)(x)\,,
\end{equation}
as a consequence of the properties \ref{item:left-right_Wigner}
and \ref{item:integral_Wigner} of the Wigner function,
and the fact that the trace takes the form 
\eqref{eq:trace_linear_case}.

\section{Conformally-coupled scalar and higher-spins}
\label{sec:HS}
Let us give two examples to show how the formalism and the action \eqref{main} can reproduce what it has to, e.g. the coupling to low-spin background fields and to higher-spin background. The latter problem was studied in $d=4$ for a coupling to a spin-three field in \cite{Beccaria:2017nco}.

\subsection{Conformally-invariant Laplacian}
As an illustration, let us show how we can recover 
the conformally-coupled scalar. This boils down to
identifying the symbol of the conformal Laplacian,
\begin{equation}
    \nabla^2 - \tfrac{n-2}{4(n-1)}\,R\,,
\end{equation}
which we can do in a couple of ways: either by
working out its quantization, or by imposing that
it transforms correctly under the above gauge transformations.

Let us start with the former. Considering that
the quantization map yields $\widehat{p}_a=-\hbar\nabla_a$,
we should consider the Ansatz $f=p^2+\alpha\,R$
for the symbol of the conformal Laplacian, where $\alpha$
is a numerical coefficient to be fixed. It is then enough
to compute the lift of this symbol, up to order $2$ in $y$,
\begin{align}
    F=\tau(p^2 + \alpha\,R)
    = p^2 + \tfrac13\,y^a y^b\,R_a{}^c{}_b{}^d\,p_c p_d
    + \alpha\,\big(R + y^a\,\nabla_a R
    + \tfrac12\,y^a y^b\,\nabla_a\nabla_b R\big) + \dots
\end{align}
as well as that 
of the scalar field $\phi$ at order $2$ in $y$
given previously in \eqref{eq:lift_phi}, 
and use
\begin{equation}\label{eq:q_lem}
    \rho(p^2)\rvert_{y=0} = \hbar^2\,\partial_y^2\,,
    \qquad\qquad
    \rho(y^a y^b p_c p_d)\rvert_{y=0}
    = \tfrac{\hbar^2}{2}\,\delta^{(a}_c\,\delta^{b)}_d\,,
\end{equation}
to finds that the quantization of the Ansatz $f$ reads
\begin{equation}
    \widehat{f}\phi = \hbar^2\,\big(\nabla^2
    +[\tfrac{\alpha}{\hbar^2} -\tfrac14]\,R\big)\phi\,,
\end{equation}
which implies
\begin{equation}
    \alpha = \frac{\hbar^2}{4(n-1)}
    \qquad\Longrightarrow\qquad
    f = p^2 + \tfrac{\hbar^2}{4(n-1)}\,R\,,
\end{equation}
upon imposing that it reproduces the conformal Laplacian.
Note that this computation also shows that, perhaps contrary
to one's intuition, the symbol of the ordinary Laplacian
is not $p^2$, but should instead be corrected by a curvature
dependent term $\tfrac{\hbar^2}{4}\,R$.

Let us now turn our attention to the symmetries
of our action, focusing on Weyl symmetry. 
Having constructed the $1$-form connection $A$
from a torsionless and metric connection, its coefficients
when expanded order by order in $y$ are tensors
built out of the vielbein and its derivatives only,
and hence have a definite behavior under Weyl transformations%
\footnote{For instance, recall that the spin-connection transforms as
$\delta_\sigma^{\Weyl} \omega^{a,b}
    = 2\,e^{[a}\,\nabla^{b]}\sigma$ under a Weyl rescaling.}
\begin{equation}
    \delta_\sigma^{\Weyl} e^a = \sigma\,e^a\,.
\end{equation}
These Weyl transformations can be realized
as gauge symmetries of $A$, by suitably choosing
the gauge parameters
$\xi_{\Weyl}, w_{\Weyl} \in \Gamma(\WeylBundle_\Base)$.
In other words, we can embed the \emph{geometric}
transformations that are Weyl rescalings,
as \emph{gauge} transformations of the system of fields
$A$, $F$ and $\Phi$ (which are affected by both types
of parameters, $\xi$ and $w$).
To explicitly find the gauge parameter
$\xi_{\Weyl}$, one needs to solve the condition
\begin{equation}
    \dR \xi_{\Weyl} + \tfrac1\hbar\,[A,\xi_{\Weyl}]_\ast
    \overset{!}{=} \delta_\sigma^{\Weyl} A\,,
\end{equation}
for $\xi_{\Weyl}$ in terms of $\sigma$.
This can be done  as before, namely order by order in $y$,
using the contracting $h$. More precisely, 
for $\xi_{\Weyl} = \sum_{k\geq1} \xi_{(k)}$
with $\xi_{(k)}$ of order $k$ in $y$ and linear in $p$,
one finds the recursion
\begin{equation}
    \xi_{(k+1)} = h\big(\nabla\xi_{(k)}
    + \tfrac1\hbar\,\sum_{l=2}^k [A_{(l)}, \xi_{(k+1-l)}]_\ast
    - \delta_\sigma^{\scriptstyle\rm Weyl} A_{(k)}\big)\,,
\end{equation}
which yields
\begin{equation}\label{eq:xi_Weyl}
    \xi_{\Weyl} = -\sigma\,y \cdot p
    - \nabla_a \sigma\,(y^a\,y \cdot p - \tfrac12\,y^2\,p^a)
    -\tfrac13\,\nabla_a \nabla_b \sigma\,\big(y^a y^b\,y \cdot p
    -\tfrac12\,y^2\,y^a\,p^b) + \dots\,,
\end{equation}
where as usual, the dots denote higher order terms in $y$.
Now we can focus on the symbol of our differential operator, 
that we assume to be of the form $p^2+\alpha\,R$
for some coefficient $\alpha$ to be fixed by requiring
that, here again, Weyl transformation can be implemented
as gauge symmetries. In other words, we want to impose
\begin{equation}
    \Big(\tfrac1\hbar\,[F, \xi_{\Weyl}]_\ast
    + \{F,w_{\Weyl}\}_\ast\Big)\big|_{y=0}
    \overset{!}{=} \delta_\sigma^{\Weyl}\big(p^2
    + \alpha\,R\big)\,, 
\end{equation}
with $F=\tau(p^2+\alpha\,R)$ its covariantly constant lift,
and where the gauge parameter $w_{\Weyl}$ is assumed 
to be proportional to the lift of the Weyl parameter $\sigma$,
i.e.
\begin{equation}
    w_{\Weyl} = \beta\,\tau(\sigma)
    \equiv \beta\,\sum_{k\geq0} \tfrac1{k!}\,
    y^{a_1} \dots y^{a_k}\,
    \nabla_{a_1} \dots \nabla_{a_k}\sigma\,,
\end{equation}
with $\beta$ a coefficient to be determined as well.
Note that at this point, the choice of $w_{\Weyl}$
is merely an educated guess: it should be covariantly constant,
and related to the Weyl parameter $\sigma$,
hence this is the simplest option---which turns out to be 
the correct one as we shall see. Using the previous formulae,
one finds on the one hand, 
\begin{equation}
    \Big(\delta_{\xi_{\Weyl},w_{\Weyl}} F\Big)\big|_{y=0}
    = 2\sigma\,(\beta+1)\,p^2
    + \tfrac{\hbar^2}{2}\,\beta\,\Box\sigma
    + 2\sigma\,\alpha\beta\,R\,,
\end{equation}
while on the other hand
\begin{equation}
    \delta^{\Weyl}_\sigma (p^2+\alpha\,R)
    = -2\alpha\,\big(\sigma\,R + (n-1)\,\Box\sigma\big)\,,
\end{equation}
which implies
\begin{equation}
    \beta=-1\,,
    \qquad\text{and}\qquad 
    \alpha = \tfrac{\hbar^2}{4(n-1)}\,,
\end{equation}
thereby fixing the symbol of the conformal Laplacian
in accordance with the previous discussion.

As a final consistency check, one can compute
the gauge transformation of the lift of the scalar $\phi$
generated by the parameter $\xi_{\Weyl}$ and $w_{\Weyl}$ 
identified previously, and recover
\begin{equation}\label{eq:var_F}
    \delta_{\xi_{\Weyl},w_{\Weyl}} \Phi\big|_{y=0} 
    = -\tfrac{n-2}2\,\sigma\,\phi\,,
\end{equation}
as expected for a conformally-coupled scalar field.

\subsection{Higher-spin background}
Let us recall that $A$ is kept purely gravitational
and background conformal higher-spin fields are placed into $F$
as an uplift of\footnote{If one wants to consider
all integer spins, a spin-one has to be included,
which is naively missing above. Alternatively,
it is possible to truncate the system to even spins only.}
\begin{align}\label{hsmetriclike}
    f & = p^2 + \tfrac{\hbar^2}{4(n-1)}\,R
    + \sum_{s>2} h^{a_1 \dots a_s}(x)\,p_{a_1} \dots p_{a_s}\,.
\end{align}
It is instructive to work out the gauge transformations
of this symbol generated by the gauge parameters
\begin{align}
    \xi & = \xi_{\text{Weyl}}
    - \tau\Big(\sum_{s>2}\xi^{a_1 \dots a_{s-1}}(x)\,
    p_{a_1} \dots p_{a_{s-1}}\Big)\,,\\
    w & = w_{\text{Weyl}}
    + \tau\Big(\sum_{s>2} \sigma^{a_1 \dots a_{s-2}}(x)\,
    p_{a_1} \dots p_{a_{s-2}}\Big)\,,
\end{align}
that is, we simply append to the gauge parameters
identified previously the covariantly constant uplift
of arbitrary monomials in $p$. Indeed, in this manner
the gauge variation of $A$ is unaffected by this new term,
\begin{equation}
    \delta_\xi A \equiv \delta_{\xi_{\Weyl}} A\,,
\end{equation}
and thus boils down to a Weyl transformation
of the gravitational sector. It does, however,
affect the gauge transformation of $f$. Computing
$\delta_{\xi,w} F\rvert_{y=0}$ and extracting
the piece of order $s>2$ in $p$, one finds
\begin{align}
    \delta_{\xi,\sigma} h^{a_1 \dots a_s}
    & = 2\,\nabla^{(a_1} \xi^{a_2 \dots a_s)}
    + 2\,\eta^{(a_1 a_2} \sigma^{a_3 \dots a_s)}
    + (s-2)\,\sigma\,h^{a_1 \dots a_s} + \dots 
\end{align}
where the dots denote curvature corrections.
The first two terms correspond to the `naive'
covariantization of the linearized gauge transformations
initially proposed by Fradkin and Tseytlin
for conformal higher-spin fields,
i.e. the flat space ones wherein partial derivatives
are replaced by covariant derivatives. The third term
tells us that the Weyl weight of a conformal higher-spin
field with spin $s$ is $s-2$, which is also in accordance
with expectations \cite{Fradkin:1985am}.%
\footnote{Note that the Weyl weight of a metric-like field
$\phi_{\mu_1 \dots \mu_s}$ is $2s-2$,
e.g. it is $2$ for metric $g_{\mu\nu}$. Its fiber version,
to which $h^{a_1 \dots a_s}$ should be compared to,
is obtained by contracting it with $s$ inverse vielbeins 
$e^\mu_a$, giving Weyl weight of $s-2$.}
This can be seen as another sign of relevance
for this framework in the problem of formulating
CHS gravity in a manifestly covariant manner.

\paragraph{Higher-spin currents.}
As a final application, we can derive the higher-spin
currents for an arbitrary curved spacetime. To do so,
let us split the previous symbol \eqref{hsmetriclike}
into that of the conformal Laplacian
and the conformal higher-spin fields,
\begin{equation}
    f = p^2 + \tfrac{\hbar^2}{4(n-1)}\,R
    + f_{hs}(x,p)\,,
    \qquad 
    f_{hs}(x,p) := \sum_{s>2} h^{a_1 \dots a_s}\,
    p_{a_1} \dots p_{a_s}\,,
\end{equation}
according to which the action obtained from $f$
is the sum of the conformally-coupled scalar
and a Noether coupling part,
\begin{equation}
    S_{\scriptstyle\rm Noether}[h,\phi]
    = \tfrac12\,\Tr_A(F_{hs} \ast \Wigner_\phi)
    = \tfrac12\,\int_\Base \dR^nx\,|e|\,\phi^*\,
    \big[\rho(F_{hs})\Phi\big]\rvert_{y=0}\,,
\end{equation}
corresponding to the contribution of the higher-spin
currents coupled to higher-spin sources/background fields
$h^{a_1 \dots a_s}$. In other words, we can identify
the higher-spin current by putting the above functional
in the form
\begin{equation}
    S_{\scriptstyle\rm Noether}[h,\phi]
    = \tfrac12\,\int_\Base \dR^nx\,|e|\
    \sum_{s>2} h^{a_1 \dots a_s}\,J_{a_1 \dots a_s}(\phi)\,,
\end{equation}
where the spin $s$ current $J_{a_1 \dots a_s}$
here is by definition bilinear in the scalar field $\phi$.

This computation involve the action of the quantization map
on the lift of $f_{hs}$, which is of arbitrary order in $p$.
As a consequence, the relevant terms to compute in this lift,
meaning those that will contribute to the final result
after applying the quantization of $F_{hs}$ to $\Phi$
and setting $y=0$, are those that are $y$-independent
or contain \emph{exactly} the same number of $y$'s and $p$'s.
Indeed, the quantization map applied to a monomial
of order $l$ in $y$ and $m$ in $p$ reads
\begin{equation}
    \rho(y^{a_1} \dots y^{a_l}\,p_{b_1} \dots p_{b_m})
    = (-\hbar)^m\,\sum_{k=0}^{\min(l,m)} \tfrac1{2^k}\,
    \tfrac{m!}{(m-k)!}\,\tfrac{l!}{k!(l-k)!}\,
    y^{(a_1} \dots y^{a_{l-k}}\,
    \delta^{a_{l+1-k}}_{(b_1} \dots \delta^{a_l)}_{b_k}\,
    \tfrac{\partial}{\partial y^{b_{k+1}}}
    \dots \tfrac{\partial}{\partial y^{b_m)}}\,,
\end{equation}
so that when setting $y=0$, only monomials with $l \leq m$,
i.e. less $y$'s than $p$'s, remain. This would be difficult
to compute for arbitrary spin $s>2$, so we will focus 
on the curvature independent part of the current.
The relevant part of the lift of $f_{hs}$
is therefore given by its `covariant Taylor series',
\begin{equation}
    F_{hs} = \sum_{k\geq0} \tfrac1{k!}\,y^{a_1} \dots y^{a_k}\,
    \nabla_{a_1} \dots \nabla_{a_k} f_{hs} + \dots\,,
\end{equation}
where the dots denote curvature corrections.
Applying the quantization map on this (partial) lift,
and evaluating the result at $y=0$, one ends up with
\begin{align}
    \rho(F_{hs})\rvert_{y=0} & = \sum_{s>2} (-\hbar)^s\,
    \sum_{k=0}^s \tfrac1{2^k} \tfrac{s!}{k!(s-k)!}\,
    \nabla_{a_1} \dots \nabla_{a_k}
        h^{a_1 \dots a_k\,a_{k+1} \dots a_s}\,
    \tfrac{\partial}{\partial y^{a_{k+1}}}
        \dots \tfrac{\partial}{\partial y^{a_s}} + \dots\,.
\end{align}
Under the same restrictions, the lift of the scalar field reads
\begin{equation}
    \Phi = \sum_{k\geq0} \tfrac1{k!}\,y^{a_1} \dots y^{a_k}\,
    \nabla_{a_1} \dots \nabla_{a_k} \phi + \dots\,,
\end{equation}
so that,
\begin{align}
    \rho(F_{hs})\Phi\rvert_{y=0} & = \sum_{s>2} (-\hbar)^s\,
    \sum_{k=0}^s \tfrac1{2^k} \tfrac{s!}{k!(s-k)!}\,
    \nabla_{a_1} \dots \nabla_{a_k}
        h^{a_1 \dots a_k\,a_{k+1} \dots a_s}
    \nabla_{a_{k+1}} \dots \nabla_{a_s} \phi
    + \dots\,,
\end{align}
again keeping only curvature independent terms.
Upon integration by parts, one finds
\begin{align}
    J_{a_1 \dots a_s} & = \big(-\tfrac{\hbar}2\big)^s\,
    \sum_{k=0}^s \tfrac{(-1)^k\,s!}{k!(s-k)!}\,
    \nabla_{(a_1} \dots \nabla_{a_k} \phi^*\,
    \nabla_{a_{k+1}} \dots \nabla_{a_{s)}} \phi 
    + \dots\,,
\end{align}
as one may have expected. This the covariantized version
of the well-known `dipole' generating function
$\phi^*(x-y)\phi(x+y)$ that yields conserved quasi-primary 
(higher-spin) currents with an admixture of descendants
in the flat space. The curvature corrections
can systematically be worked out, see \cite{Beccaria:2017nco}
for the spin-three example in the bottom-up approach.
However, it is clear that the higher the spin
the more non-linearities in the Riemann tensor $R$
and its derivatives will enter. Therefore, eq. \eqref{main}
seems to be the most compact way of writing the coupling
of the free scalar field to a higher-spin background.

\paragraph{First order correction in curvature.}
If we focus on the spin-$3$ case, then we only need
to compute the lift of $f_{s=3} = h^{abc}\,p_a p_b p_c$
to order $3$. Pushing the computation of the lift
of any symbol $f(x,p)$ presented in \eqref{eq:lift_y2}
to the next order, thanks to the recursion \eqref{eq:rec_F},
yields
\begin{align}
    \tau(f) & = \Big(1 + y^a \nabla_a
    + \tfrac12\,y^a y^b\,\big[\nabla_a \nabla_b 
    + \tfrac13\,R_{da}{}^c{}_b\,p_c\,
    \tfrac{\partial}{\partial p_d}\big] \\
    & \hspace{50pt} + \tfrac16\,y^a y^b y^c\,
    \big[\nabla_a \nabla_b \nabla_c
    + \tfrac12\,\nabla_a R_{db}{}^e{}_c\,
    p_e\,\tfrac{\partial}{\partial p_d}
    + R_{da}{}^e{}_b\,p_e\,
    \tfrac{\partial}{\partial p_d}\,\nabla_c\big] \\
    & \hspace{100pt} + \tfrac{\hbar^2}{12}\,
    y^a\,\big[R_{ab}{}^d{}_c\,
    \tfrac{\partial^2}{\partial p_b \partial p_c}\nabla_d
    - \tfrac14\,\nabla_b R_{ac}{}^e{}_d\,p_e\,
    \tfrac{\partial^3}{\partial p_b \partial p_c \partial p_d}\big]
    + \dots \Big)\,f\,,
\end{align}
and applying it to $f_{s=3}$, one finds that
$F_{s=3} = \tau(f_{s=3})$ is given by
\begin{align}
    F_{s=3} & = \Big(h^{abc} + y^i\,\nabla_i h^{abc}
    + \tfrac12\,y^i y^j\,\big[\nabla_i \nabla_j h^{abc}
    + R_{di}{}^a{}_j h^{bcd}\big] \\
    & \hspace{50pt} + \tfrac16\,y^i y^j y^k\,
    \big[\nabla_i \nabla_j \nabla_k h^{abc} 
    + \tfrac32\,\nabla_i R_{dj}{}^a{}_k\,h^{bcd}
    + 3\,R_{di}{}^a{}_j\,\nabla_k h^{bcd}\big] + \dots\Big)\,p_a p_b p_c \\
    & \hspace{100pt} + \tfrac{\hbar^2}{2}\,y^i\,
    \big[R_{ib}{}^d{}_c\,\nabla_d h^{abc} 
    - \tfrac14\,\nabla_b R_{ic}{}^a{}_d\,h^{bcd}\big]\,p_a
    + \dots\,,
\end{align}
where the dots denote terms of order $4$ or higher,
and its quantization evaluated at $y=0$ reads
\begin{align}
    -\tfrac1{\hbar^3}\,\rho(F_{s=3})\rvert_{y=0}
    & = h^{abc}\,\tfrac{\partial^3}{\partial y^a \partial y^b \partial y^c}
    + \tfrac32\,\nabla_a h^{abc}\,
    \tfrac{\partial^2}{\partial y^b \partial y^c}
    + \tfrac14\,\big[3\,\nabla_a \nabla_b h^{abc}
    -R_{ab}\,h^{abc}\big]\,\tfrac{\partial}{\partial y^c} \\
    & \hspace{100pt}
    + \tfrac18\,\big[\nabla_a \nabla_b \nabla_c h^{abc}
    - \nabla_a R_{bc}\,h^{abc} - R_{ab}\,\nabla_c h^{abc}\big]\,.
\end{align}
Similarly, we can use \eqref{eq:rec_phi} to compute
the lift of the scalar field $\phi$ to order $3$,
\begin{equation}
    \Phi = \phi + y^a\nabla_a\phi
    + \tfrac12\,y^a y^b\,\big(\nabla_a\nabla_b
    - \tfrac16\,R_{ab}\big)\phi
    + \tfrac16\,y^a y^b y^c\,\big(\nabla_a \nabla_b \nabla_c
    -\tfrac14\,\nabla_a R_{bc}
    -\tfrac12\,R_{ab}\,\nabla_c\big)\,\phi + \dots\,,
\end{equation}
so that one finds
\begin{align}
    -\tfrac1{\hbar^3}\,\rho(F_{s=3})\Phi\rvert_{y=0}
    & = h^{abc}\,\big(\nabla_a \nabla_b \nabla_c
    -\tfrac38\,\nabla_a R_{bc}
    -\tfrac34\,R_{ab}\,\nabla_c\big)\,\phi
    + \tfrac32\,\nabla_c h^{abc}\,
    \big(\nabla_a\nabla_b- \tfrac14\,R_{ab}\big)\,\phi \\
    & \quad + \tfrac34\,\nabla_a \nabla_b h^{abc}\,\nabla_c\phi
    + \tfrac18\,\nabla_a \nabla_b \nabla_c h^{abc}\,\phi\,.
    \nonumber
\end{align}
which leads to the following expression
\begin{align}
    J_{abc} & = -\tfrac{\hbar^3}{8}\,
    \Big(\nabla_{(a} \nabla_b \nabla_{c)} \phi\,\phi^*
    - 3\,\nabla_{(a} \nabla_b \phi\,\nabla_{c)} \phi^*
    + 3\,\nabla_{(a} \phi\,\nabla_b \nabla_{c)} \phi^*
    - \phi\,\nabla_{(a} \nabla_b \nabla_{c)} \phi^* \\
    & \hspace{250pt} -3\,R_{(ab}\,\nabla_{c)} \phi\,\phi^*
    + 3\,R_{(ab}\,\phi\,\nabla_{c)} \phi^*\Big)\,,
\end{align}
for the spin-$3$ current. More generally, the currents
up to first order in the curvature tensor (and its derivatives)
are obtained from the generating function
\begin{equation}
    {\cal J}(x|u) = e^{-\frac\hbar2\,u \cdot[\nabla_1-\nabla_2]}\,
    \Big(1 - \tfrac{\hbar^2}8\,
    {\rm sinhc}(\tfrac\hbar4\,u \cdot \nabla_3)\,
    R_{ab}(x_3)\,u^a u^b + {\cal O}(R^2)\Big)\,\phi(x_1)\,\phi^*(x_2)\big|_{x_i=x}\,,
\end{equation}
where
\begin{equation}
    {\cal J}(x|u) := \sum_{s \geq 0} \tfrac{(-\hbar)^s}{2^s\,s!}\,
    J_{a_1 \dots a_s}(x)\,u^{a_1} \dots u^{a_s}\,,
    \qquad 
    {\rm sinhc}(z) := \frac{\sinh(z)}{z}\,,
\end{equation}
and $\nabla_i$ denotes the covariant derivative
with respect to $x_i$ (see Appendix \ref{app:curvature}
for the derivation of this formula).
Let us remark here that, in order to make contact 
with, say the computation of \cite{Beccaria:2017nco}
for the spin-three current or even the standard computation
of the energy-momentum tensor, one should find the correct field
redefinition bringing the components of monomials in $p$
into the appropriate field frame (combination of conformal higher spin fields and derivatives thereof),
see e.g. \cite[App. F]{Segal:2002gd} and \cite{Bekaert:2010ky}
for an instance of the same issue around a flat background.

\section{Discussion}
\label{sec:discu}
The present paper is a natural continuation of the quest
to covariantize the construction of conformal higher-spin 
gravities started in \cite{Basile:2022nou}.
Now, both the action for conformal higher-spin gravity
$S_{CHS}[h_s]$ and the coupling of the scalar matter
to the higher-spin background,
$\bra\Phi\widehat{H}[\phi_s]\ket\Phi$,
can be written in a covariant way. The result completes
the study initiated in \cite{Beccaria:2017nco},
where the mixing between covariant spin-three
and spin-one currents that couple to background fields
have been discussed in $n=4$. In addition, one can consider
the matter coupled conformal higher-spin gravity,
see \cite{Joung:2015eny} for some amplitudes
in this theory over flat background. Note, however,
that while the scalar matter can be coupled
to a higher-spin background for any $n$ the conformal anomaly 
recipe gives $S_{CHS}[h_s]$ only for $n$ even.

The results open up the possibility of considering more general
matter fields in the relevant higher-spin background,
such as the higher-derivative scalar fields (also known as
higher order singletons \cite{Bekaert:2013zya}), 
or spinor (and its higher-derivative counterpart),
see \cite{Grigoriev:2018wrx}.
The latter would in principle require the use
of the supersymmetry version of the FFS cocycle, 
i.e. the representative of the cohomology class
of the Clifford--Weyl algebra dual to the unique
Hochschild homology class of the same algebra 
\cite{Engeli:2008}.

Another possible application of the results
is to conformally-invariant differential operators.
Conformal geometry (in the sense of gauge symmetries
realized by diffeomorphisms and Weyl transformations)
is a part of the higher-spin system. As we showed,
one can derive the conformal Laplacian as a particular instance 
of the scalar field coupled to the conformal gravity 
background. Generalizations such as Paneitz \cite{Paneitz:2008} 
or Fradkin--Tseytlin \cite{Fradkin:1982xc} operators
and GJMS operators \cite{GJMS} can also be recovered
by considering $F = (p^2)^k + \dots$ 
that would lead to operators of type $(\nabla^2)^k+\dots$,
i.e. starting with the $k$th power of the Laplacian,
and corrected by curvature terms.

It would also be interesting to apply the deformation 
quantization techniques to the self-dual conformal higher-spin 
gravity \cite{Hahnel:2016ihf,Adamo:2016ple} that is natural
to formulate on twistor space. Here, the underlying space
$\mathbb{CP}^3$ is already symplectic. The twistor description
of low-spin fields, $s=1,2$, requires usual (holomorphic) 
connections and vector-valued one-forms, which can be understood 
as differential operators of zeroth and first order.
An extension to higher-spin calls for differential operators
of arbitrary order, i.e. to the quantization
of the cotangent bundle again
(see also \cite{Bekaert:2021sfc} for additional discussions
of the quantization of the cotangent in relation with
the definition of higher-spin diffeomorphisms).

Let us also note that the present paper bridges a gap
in the phase space approach to quantum mechanics.
Indeed, one can attempt to extend the Fedosov construction
to accommodate all the usual ingredients required
in quantum mechanics. The trace is, obviously, given by
the Feigin--Felder--Shoikhet cocycle; wave functions
can be understood as covariantly constant elements
in the Fock representation obtained via the quantization map. 
Wigner function takes exactly the same form as in the flat space, 
but in the fiber. The basic ingredients above do not rely
on the phase space being a cotangent bundle and should extend to 
arbitrary symplectic manifolds (a polarization is needed
to define the Fock space). This seems to depart
from the usual approach of symbol calculus on curved background, 
e.g. \cite{Hormander:1985, Duistermaat:1994, Widom:1980, Safarov:1997, Pflaum:1998, Pflaum:1999,  Dereziski:2018}
and references therein.

Finally, it would be interesting to construct
the $3d$ matter-coupled conformal higher-spin gravity,
where the `dynamics' of conformal higher-spin fields
is given by the Chern--Simons action
(as there is no conformal anomaly in $3d$).
Such a theory, namely the one based on fermionic matter,
can be seen to exist with the help of the argument
based on the parity anomaly \cite{Grigoriev:2019xmp}
(see e.g. \cite{Niemi:1983rq, Redlich:1983dv, Redlich:1983kn}
for original papers on the derivation of Chern--Simons
theory from the parity anomaly and \cite{Bonora:2016ida}
for the spin-three case). An alternative idea
along the AdS/CFT correspondence lines
was recently explored in \cite{Diaz:2024kpr, Diaz:2024iuz}.

\section*{Acknowledgements}
We are grateful to Pierre Bieliavsky and Maxim Grigoriev for useful discussions. This project has received funding from
the European Research Council (ERC)
under the European Union’s Horizon 2020 research
and innovation programme (grant agreement
No. 101002551). E.S. is a Research Associate
of the Fund for Scientific Research - FNRS, Belgium.
The work of T.B. was also supported by the European Union’s 
Horizon 2020 research and innovation programme
under the Marie Sk\l{}odowska Curie grant agreement No. 101034383.

\appendix
\section{A brief review of Weyl calculus}
\label{app:Weyl_calculus}
Let us give a brief summary of the definition
and construction of the Wigner function in flat space
(following e.g. the textbook \cite{deGosson:2006},
or the papers \cite{Segal:2002gd, Bekaert:2008xfa, Bekaert:2009ud, Bekaert:2010ky}).

\paragraph{Quantization map in flat space.}
The deformation quantization of $\R^{2n} \cong T^*\R^n$,
amounts to defining an isomorphism
\begin{equation}
    \begin{tikzcd}
        \Functions(T^*\R^n) \ar[r, "\sim"]
        & {\cal D}(\R^n)\,,
    \end{tikzcd}
\end{equation}
where ${\cal D}(\R^n)$ stands for the space
of differential operators on $\R^n$. This map
is referred to as a `quantization map' since,
as we will recall shortly, it allows one to define
a star-product on the algebra of functions $T^*\R^n$,
and hence a quantization thereof. To do so, 
we can take advantage of the Fourier transform
in flat space, that we denote by
\begin{equation}
    ({\cal F}f)(u,v) := \int_{\mathbb R^{2n}}
    \tfrac{{\rm d}^nx\,{\rm d}^np}{(2\pi\hbar)^n}\,
    f(x,p)\,e^{-\frac{i}\hbar\,(x \cdot u + p \cdot v)}\,,
\end{equation}
for a symbol $f(x,p)$. Given a choice of quantization 
for the phase space coordinates $x^\mu \to \hat x^\mu$
and $p_\mu \to \hat p_\mu$, where hatted symbols
denote the corresponding operator, we want to associate

Schematically, we want to write something like
``$\widehat f(\hat x, \hat p)
\sim f(x,p)\,\delta(x-\hat x)\,\delta(p-\hat p)$'',
where $f(x,p)$ is the symbol of the operator $\widehat f$.
This sketchy formula can be given a precise sense,
using the Fourier representation of the Dirac distribution, 
leading to
\begin{equation}
    \widehat f(\hat x, \hat p) = \int_{\mathbb R^{2n}}
    \tfrac{{\rm d}^nu\,{\rm d}^nv}{(2\pi\hbar)^n}\,
    ({\cal F}f)(u,v)\,
    e^{\frac{i}\hbar\,(u \cdot \hat x + v \cdot \hat p)}\,,
\end{equation}
and which is called the \emph{Weyl ordering} of operators.
Note that the exponential operator can be re-written as
\begin{equation}\label{eq:BCH}
    \exp\big(\tfrac{i}{\hbar}\,(u \cdot \hat x
                                + v \cdot \hat p)\big)
    = e^{\frac{i}{2\hbar}\,u \cdot v}\,
    \exp\big(\tfrac{i}{\hbar}\,u \cdot \hat x\big)\,
    \exp\big(\tfrac{i}{\hbar}\,v \cdot \hat p\big)\,,
\end{equation}
since we assume
$[\hat x^\mu, \hat p_\nu]=i\hbar\,\delta^\mu_\nu$.
Choosing the usual coordinate representation,
\begin{equation}
    \hat x^\mu = x^\mu\,,
    \qquad 
    \hat p_\mu = -i\hbar\,\partial_\mu\,,
\end{equation}
the action of this operator on a wave function $\varphi(x)$
is given by
\begin{align}
    (\widehat f\,\varphi)(x)
    & = \int \tfrac{\dR^nu\,\dR^nv}{(2\pi\hbar)^n}\,
    ({\cal F}f)(u,v)\,e^{\frac{i}{2\hbar}\,u \cdot v}\,
    e^{\frac{i}\hbar\,u \cdot x}\,\varphi(x+v) \\
    & = \int \tfrac{\dR^nu\,\dR^nv}{(2\pi\hbar)^n}\,
    \tfrac{{\rm d}^nx'\,{\rm d}^np}{(2\pi\hbar)^n}\,
    f(x',p)\,e^{-\frac{i}\hbar\,p \cdot v}\,
    e^{\frac{i}\hbar\,u \cdot (x-x'+\frac{v}2)}\,
    \varphi(x+v) \\
    & = \int \tfrac{\dR^nv\,\dR^np}{(2\pi\hbar)^n}\,
    f(\tfrac{x+v}2,p)\,e^{\frac{i}\hbar\,p \cdot (x-v)}\,
    \varphi(v)\,,
\end{align}
where the first equation is obtained using
\eqref{eq:BCH} and the action of the translation operator,
the second line is merely the definition
of the Fourier transform, and the last one is the result
of integrating over $u$, which gives a Dirac distribution,
and then evaluating it by integrating over $x'$.
Upon Taylor expanding $f$ and integrating by part,
one can put this formula into an operatorial form
\begin{equation}\label{eq:exp_quantization}
    (\widehat f\varphi)(x) = f(x,p)\,
    \exp\Big(\!\!-\!i\hbar\,
    \tfrac{\overleftarrow{\partial}}{\partial p} \cdot
    \big[\tfrac12\,\tfrac{\overleftarrow{\partial}}{\partial x}
    + \tfrac{\overrightarrow{\partial}}{\partial x}\big]\Big)\,
    \varphi(x)\big|_{p=0}\,.
\end{equation}

The Moyal--Weyl star-product can be recovered
from the composition of the two operators associated with
two symbols via the above symbol, or quantization, map.
More precisely, it can be defined as the symbol
of the composition of the quantization of two symbols, i.e.
\begin{equation}\label{eq:quantization=rep}
    \widehat f \circ \widehat g = \widehat{f \star g}\,.
\end{equation}
To do so, let us start by recalling
that the action of a symbol $f$ given above
exhibits the \emph{kernel} of that associated operator, 
namely
\begin{equation}
    (\widehat f\,\phi)(x) = \int_{\R^n} \dR^nq\,
    K_f(x,q)\,\phi(q)\,,
    \qquad\text{with}\qquad
    K_f(x,q) := \int_{\R^n} \tfrac{\dR^np}{(2\pi\hbar)^n}\,
    f(\tfrac{x+q}2,p)\,e^{\frac{i}\hbar\,p \cdot (x-q)}\,.
\end{equation}
The symbol of the operator $\widehat f$ can be extract back
from its kernel, via its inverse transform
\begin{equation}
    f(x,p) = \int_{\R^n} \dR^nq\,
    K_f(x+\tfrac12\,q,x-\tfrac12\,q)\,
    e^{-\frac{i}\hbar\,p \cdot q}\,,
\end{equation}
and therefore, using this together with the fact
that the integral kernel of the composition
of two operators is given by
\begin{equation}
    K_{\widehat f \circ \widehat g}(x,x')
    = \int_{\R^n} \dR^nq\,
    K_{\widehat f\,}(x,q)\,K_{\widehat g\,}(q,x')\,,
\end{equation}
one ends up with
\begin{equation}
    \big(f \star g\big)(x,p) = \tfrac1{(\pi\hbar)^{2n}}
    \int \dR^nv_1\,\dR^nv_2\,\dR^nw_1\,\dR^nw_2\
    e^{\frac{2i}\hbar\,(v_1 \cdot w_2 - v_2 \cdot w_1)}\,
    f(x+v_1,p+w_1)\,g(x+v_2,p+w_2)\,.
\end{equation}
Upon Taylor expanding the two functions $f$ and $g$
around $(x,p)$, and integrating by part, 
one 
\begin{equation}\label{eq:exp_star}
    \big(f \star g\big)(x,p) = f(x,p)\,
    \exp\Big(\tfrac{i\hbar}2\,
    \big[\tfrac{\overleftarrow{\partial}}{\partial x}
    \cdot \tfrac{\overrightarrow{\partial}}{\partial p}
    -\tfrac{\overleftarrow{\partial}}{\partial p}
    \cdot \tfrac{\overrightarrow{\partial}}{\partial x}\big]\Big)\,g(x,p)\,,
\end{equation}
Note that the Moyal--Weyl star-product is \emph{Hermitian},
meaning that it satisfies
\begin{equation}
    (f \star g)^* = g^* \star f^*\,,
\end{equation}
where $(-)^*$ denotes the complex conjugation,
i.e. the latter is an anti-involution of the Weyl algebra.

One can think of the quantization map as providing
a \emph{representation} of the Weyl algebra: identifying
the latter as the subalgebra of \emph{polynomial}
functions on $T^*\R^n$, wave functions which are nothing but 
functions on $\R^n$, the base of the cotangent bundle $T^*\R^n$,
are acted upon by the former via the quantization map.
This subspace can be thought of as a Fock space,
which carries a representation of the Weyl algebra
as can be seen from the defining relation 
\eqref{eq:quantization=rep}.

The integration over the cotangent bundle
$T^*\R^n$ defines a trace over the space of symbols,
at least those which are compactly supported
or vanish at infinity sufficiently fast.
Indeed, in this case one finds
\begin{equation}
    \Tr(f \star g) = \int_{T^*\R^n} \dR^n x\,\dR^n p\
    (f \star g)(x,p) = \int_{T^*\R^n} \dR^n x\,\dR^n p\
    f(x,p)\,g(x,p) = \Tr(g \star f)\,,
\end{equation}
for any symbols $f$ and $g$, since all higher order terms
in the star product are total derivatives on $T^*\R^n$,
and hence can be ignored for the aforementioned
suitable class of symbols.

\paragraph{Wigner function in flat space.}
Having worked out how to translate the action
and composition of differential operators in terms
of their symbol, as well as their trace, we can now
turn our attention to the computation of matrix elements
for these operators, expressing the transition probability
from one state to another. Since the latter can be expressed as
\begin{equation}\label{eq:transition}
    \bra\psi\widehat{H}\ket\phi
        = \Tr(\widehat{H} \circ \ket\phi\!\bra\psi)\,,
\end{equation}
we have everything we need to derive such quantities
using symbols, provided that we know that of the projector
$\ket\phi\!\bra\psi$. In light of the relation
between the symbol of an operator and its integral kernel,
we may first focus on that of the projector.
This integral kernel is easily computed,
\begin{equation}
    \big(\ket{\phi}\!\bra{\psi}\ket{\varphi}\big)(x)
    \overset{!}{=} \phi(x)\,\int_{\R^n} \dR^nq\,
    \psi^*(q)\,\varphi(q)
    \qquad\Longrightarrow\qquad 
    K_{\ket{\phi}\!\bra{\psi}}(x,q)
    \equiv \phi(x)\,\psi^*(q)\,,
\end{equation}
which leads to 
\begin{equation}
    W[\phi,\psi](x,p) = \int_{\R^n} \dR^nq\
    \phi(x+\tfrac{q}2)\,\psi^*(x-\tfrac{q}2)\,
    e^{-\frac{i}\hbar\,p \cdot q}\,,
\end{equation}
for its symbol. It obey the following useful properties
\begin{equation}\label{eq:prop_Wigner}
    \xi \star W[\phi,\psi]
        = W[\widehat\xi\,\phi,\psi]\,,
    \qquad 
    W[\phi,\psi] \star \xi^\dagger
        = W[\phi,\widehat\xi\,\psi]\,,
\end{equation}
in accordance with the fact that it is the symbol
of the projector $\ket\phi\!\bra\psi$, and
\begin{equation}\label{eq:integral_Wigner}
    \int_{\R^n} \dR^np\ W[\phi,\psi](x,p)
    = \phi(x)\,\psi^*(x)\,.
\end{equation}
Now we can replace the right hand side of \eqref{eq:transition} with its symbol counterpart, leading to
\begin{equation}
    \Tr(H \star W[\phi,\psi]) = \Tr(W[\widehat{H}\phi,\psi])
    = \int_{T^*\R^n} \dR^nx\,\dR^np\ W[\widehat{H}\phi,\psi]
    = \int_{\R^n} \dR^nx\ \psi^*(x)\,(\widehat{H}\phi)(x)\,,
\end{equation}
upon using the previously listed properties of $W[\phi,\psi]$,
thereby reproducing the expected result for the quantity
$\bra\psi\widehat{H}\ket\phi$ from a quantum mechanical
point of view. The \emph{Wigner function} $W_\phi$
associated with a wave function $\phi$ is the symbol 
of the projector $\ket\phi\!\bra\phi$, i.e. 
\begin{equation}
    W_\phi(x,p) := W[\phi,\phi](x,p)
    \equiv \int_{\R^n} \dR^nq\ e^{-\frac{i}\hbar\,q \cdot p}\,
    \phi(x+\tfrac{q}2)\,\phi(x-\tfrac{q}2)\,,
\end{equation}
whose integral over $p$ is nothing but the probability
density defined by $\phi$.

To conclude this appendix, let us prove the identity
\eqref{eq:integral_Wigner} and a small variation on it
(the intertwining property \eqref{eq:prop_Wigner}
can be proved by direct computation using the integral formulae 
for the star-product and the quantization map),
by expressing the Wigner function in terms of star-product.
To achieve this, recall that the star-product
of a phase factor $e^{\frac{i}\hbar\,q \cdot p}$,
where $q$ is a fixed parameter, with any symbol $f(x,p)$
yields
\begin{equation}\label{eq:translation}
    e^{\frac{i}\hbar\,q \cdot p} \star f(x,p)
    = e^{\frac{i}\hbar\,q \cdot p} f(x+\tfrac{q}2,p)\,,
    \qquad 
    f(x,p) \star e^{\frac{i}\hbar\,q \cdot p}
    = e^{\frac{i}\hbar\,q \cdot p} f(x-\tfrac{q}2,p)\,,
\end{equation}
i.e. it implements translations in $x$ up to a phase.%
\footnote{To be more precise, the action of translation
on elements depending on $x$ \emph{only} is generated by 
\[
    e^{\frac{i}\hbar\,q \cdot p} \star \phi(x)
    \star e^{-\frac{i}\hbar\,q \cdot p} = \phi(x+q)\,,
\]
which can be recovered from the formulae \eqref{eq:translation}.}
Integrating these formulae over $q$ yields
\begin{equation}
    \big(f \star \delta_p\big)(x,p)
    = \int_{\R^n} \dR^nq\,e^{-\frac{i}\hbar\,q \cdot p}\,
    f(x+\tfrac{q}2,p)\,, 
    \qquad 
    \big(\delta_p \star f\big)(x,p)
    = \int_{\R^n} \dR^nq\,e^{-\frac{i}\hbar\,q \cdot p}\,
    f(x-\tfrac{q}2,p)\,,
\end{equation}
where $\delta_p$ is the Dirac distribution
in the space of momenta $p_a$. With these simple identities
at hand, one finds
\begin{align}
    \phi \star \delta_p \star \psi^*
    & = \int_{\R^n} \dR^nq\, \phi(x)
    \star \big[e^{-\frac{i}\hbar\,q \cdot p}
    \psi^*(x-\tfrac12\,q)\big] \\
    & = \int_{\R^n} \dR^nq\,\big[\phi(x)
    \star e^{-\frac{i}\hbar\, q \cdot p}]\,
    \psi^*(x-\tfrac12\,q) \\
    & = \int_{\R^n} \dR^nq\
    \phi(x+\tfrac{q}2)\,\psi^*(x-\tfrac{q}2)\,
    e^{-\frac{i}\hbar\,p \cdot q} = W[\phi,\psi]\,,
\end{align}
where to pass from the first to the second line,
one should notice that since $\phi$ only depends on $x$,
its star-product with any other Weyl algebra element
will produce only derivatives with respect to $p$ 
on the latter.

Now this expression makes it  relatively easy
to evaluate the integral over momenta of the Wigner function
and its derivatives with respect to $p$. Indeed,
since the only term of this star-product that depends
on momenta is the Dirac distribution, the result is of the form
\begin{equation}
    \phi \star \delta_p \star \psi^*
    \sim \sum_{k,l\geq0} \partial_x^k\phi \times
    \partial_p^{k+l} \delta(p) \times \partial_x^l \psi^*\,,
\end{equation}
so that the integral over $p$ schematically reads
\begin{equation}
    \int \dR^np\ W[\phi,\psi] \sim \sum_{k,l\geq0}
    \int \dR^np\ \partial_p^{k+l} \delta(p)
    \times (\partial_x^k\phi\,\partial_x^l\psi^*)\,,
\end{equation}
which identically vanishes for $k+l>0$ since both
$\phi$ and $\psi$ do not depend on $p$, and yields
\eqref{eq:integral_Wigner} for $k=0=l$. On top of that,
since taking partial derivative with respect to $x$ or $p$
commutes with the star-product, the derivatives
of the Wigner function with respect to $p$ are of the form
$\partial_p^k W[\phi,\phi]
    \sim \phi \star \partial^k_p\delta_p \star \psi^*$,
and hence the same argument shows that the integral
over the momenta identically vanishes, 
\begin{equation}\label{eq:der_Wigner}
    \int_{\R^n} \dR^np\ \tfrac{\partial^k}{\partial p_{a_1} \dots \partial p_{a_k}}\,W[\phi,\psi] = 0\,,
    \qquad \forall k>0\,.
\end{equation}

\section{More on Weyl transformations}
\label{app:Weyl_transfo}
In this appendix, we provide more details concerning
the computation of the gauge variation of the symbol
$p^2 + \alpha\,R$. For convenience, let us introduce 
the tensor
\begin{equation}
    \mathcal{P}_{ab}{}^{cd} := \delta_{(a}^c \delta_{b)}^d
    - \tfrac12\,\eta_{ab}\,\eta^{cd}\,,
\end{equation}
with which the gauge parameter $\xi_{\Weyl}$,
identified in \eqref{eq:xi_Weyl} as the one generating
Weyl transformations for the components of
the $1$-form connection $A$, is given by
\begin{equation}
    \xi_{\Weyl} = -\sigma\,y \cdot p
    - \mathcal{P}_{bc}{}^{ad}\,y^b y^c p_d\,\nabla_a \sigma
    -\tfrac13\,y^{(a}\,\mathcal{P}_{cd}{}^{b)e} y^c y^d p_e\,
    \nabla_a \nabla_b \sigma + \dots\,,
\end{equation}
plus terms of order $3$ and higher in $y$, but all linear in $p$.
In order to compute the gauge transformation
of $F = \tau(p^2+\alpha\,R)$ generated by 
$\xi_{\Weyl}$, and $w_{\Weyl}$ given by
\begin{equation}
    w_{\Weyl} = \beta\,\tau(\sigma)
    = \beta\,\big(1 + y^a\,\nabla_a
    + \tfrac12\,\nabla_a \nabla_b + \dots\big)\,\sigma\,,
\end{equation}
one needs to compute the star-product between elements
of the Weyl algebra which are at most quadratic in $p$.
For our purpose, it will be enough to compute
neglecting terms with less, or as many, $y$'s than $p$'s.
We therefore only need the lift of $p^2$ and $R$
up to order $2$ in $y$,
\begin{equation}
    \tau(p^2) = p^2 + \tfrac13\,y^a y^b\,
    R_a{}^c{}_b{}^d\,p_c p_d + \dots\,,
    \qquad 
    \tau(R) = R + y^a\,\nabla_a R
    + \tfrac12\,y^a y^b\,\nabla_a \nabla_b R + \dots\,,
\end{equation}
which yields
\begin{align}
    \tfrac1\hbar\,\big[p^2, \xi_{\Weyl}\big]_\ast
    & = 2\sigma\,p^2 + 4\,\mathcal{P}_{bc}{}^{da}\,
    y^b\,p^c p_d\,\nabla_a \sigma
    + \tfrac23\,y^c y^d\,\mathcal{P}_{cd}{}^{a(\bullet}\,p^{b)}\,
    p_a\,\nabla_b \nabla_\bullet \sigma \\ & \hspace{150pt}
    + \tfrac43\,y^{(b}\,\mathcal{P}_{de}{}^{c)a}\,
    y^e\,p^d p_a\,\nabla_b \nabla_c \sigma + \dots \\
    & = 2\,p^2\,\big(\sigma + y^a\,\nabla_a \sigma
    + \tfrac13\,y^a y^b\,\nabla_a \nabla_b \sigma\big)
    + \tfrac23\,(y \cdot p\,y^a - \tfrac12\,y^2\,p^a)\,p^b\,
    \nabla_a \nabla_b \sigma + \dots\,, 
\end{align}
while the commutator of $\xi_{\Weyl}$ with other terms
in the lift of $p^2$ or $R$ do not contribute terms
with less $y$'s than $p$'s, and
\begin{align}
    \big\{\tau(p^2), w_{\Weyl}\big\}_\ast
    & = \Big(2\,\big[p^2 + \tfrac13\,R_{a}{}^{c}{}_{b}{}^{d}\,
    y^a y^b\,p_c p_d\big] + \tfrac{\hbar^2}2\,\big[\eta^{ab}
    + \tfrac13\,R_{a}{}^{c}{}_{b}{}^{d}\,y^a y^b\big]\,
    \tfrac{\partial^2}{\partial y^a \partial y^b}\big)
    w_{\Weyl} + \dots \\
    & = 2\beta\,p^2\,\big(\sigma + y^a\,\nabla_a \sigma
    + \tfrac12\,y^a y^b\,\nabla_a \nabla_b \sigma\big)
    + \tfrac{2\beta}3\,\sigma\,R_{a}{}^{c}{}_{b}{}^{d}\,
    y^a y^b\,p_c p_d + \beta\,\tfrac{\hbar^2}2\,\Box\sigma
    + \dots \\
    \big\{\tau(R), w_{\Weyl}\big\}_\ast
    & = 2\,\tau(R)\,w_{\Weyl} = 2\beta\,\sigma\,R + \dots 
\end{align}
where again the dots denote terms of order $3$
or higher in $y$.
Putting everything together, we end up with
\begin{align}
    \delta_{\xi_{\Weyl},w_{\Weyl}} F
    & = 2\sigma\,\big[(\beta+1)\,p^2 + \alpha\beta\,R\big]
        + \beta\,\tfrac{\hbar^2}2\,\Box\sigma 
        + 2(\beta+1)\,p^2\,y^a\,\nabla_a \sigma \\
    & \qquad + y^a y^b\,p_c p_d\,\big([\beta+\tfrac23]\,
    \eta^{cd}\,\delta^\times_a \delta^\bullet_b
    + \tfrac23\,\eta^{\times c}\,\delta^d_a \delta^\bullet_b
    -\tfrac13\,\eta_{ab}\,\eta^{\times c}\eta^{\bullet d}\big)\,
    \nabla_\times \nabla_\bullet \sigma + \dots
\end{align}
whose value at $y=0$, which we gave earlier in \eqref{eq:var_F},
can be compared to the Weyl variation of $p^2+\alpha\,R$
and imposing that the two agree implies
\begin{equation}
    \alpha = \tfrac{\hbar^2}{4(n-1)}\,,
    \qquad 
    \beta = -1\,.
\end{equation}
From now on, we will fix these values, 
and will denote the gauge transformations
generated by $\xi_{\Weyl}$ and $w_{\Weyl}$ 
with the same symbol as for a Weyl transformation
generated by $\sigma$,
\begin{equation}
    \delta_{\xi_{\Weyl},w_{\Weyl}} \equiv \delta_\sigma\,,
\end{equation}
since the two agree with the aforementioned values
of $\alpha$ and $\beta$.
As a final cross-check, let us compute
the Weyl transformation of the equation of motion
\begin{equation}
    \widehat{f}\phi
    = \big(\Box - \tfrac{n-2}{4(n-1)}\,R\big)\,\phi\,,
\end{equation}
which, in our formalism, is obtained by evaluating
\begin{equation}
    \delta_\sigma\big(\rho(F)\,\Phi\big)
    = \rho(\delta_\sigma F)\,\Phi
    + \rho(F)\,\delta_\sigma\Phi\,,
\end{equation}
at $y=0$, where recall that $\Phi=\tau(\phi)$
is the lift of $\phi$ as a flat section of the Fock bundle,
whose first order in $y$ are given in \eqref{eq:lift_phi}.
To compute the first term, we only need to use
the simple quantization formula \eqref{eq:q_lem},
to find
\begin{equation}\label{eq:interm1}
    \rho(\delta_\sigma F)\,\Phi\rvert_{y=0}
    = \hbar^2\,\tfrac{n-4}6\,\big(\Box\sigma
    + \tfrac1{n-1}\,R\,\sigma\big)\,\phi\,.
\end{equation}
To compute the second term, we should also use
\begin{equation}
    \rho(F)\rvert_{y=0} = \hbar^2\,\eta^{ab}\,
    \tfrac{\partial^2}{\partial y^a \partial y^b}
    - \tfrac{\hbar^2}{12}\,\tfrac{n-4}{n-1}\,R\,,
\end{equation}
as well as
\begin{equation}
    \tfrac1\hbar\,\rho(\xi_{\Weyl})
    = -\big(\sigma + y^a\,\nabla_a \sigma\big)\,
    \big(y \cdot \tfrac{\partial}{\partial y} + \tfrac{n}2\big)
    + \tfrac12\,y^2\,\nabla^a\,\tfrac{\partial}{\partial y^a}
    - \tfrac{n+1}3\,y^a y^b\,\nabla_a \nabla_b \sigma
    + \tfrac16\,y^2\,\Box\sigma + \dots\,,
\end{equation}
and
\begin{equation}
    \rho(w_{\Weyl})= -\big(1 + y^a\,\nabla_a
    + \tfrac12\,[\nabla_a \nabla_b - \tfrac16\,R_{ab}]
    + \dots\big)\sigma\,.
\end{equation}
Acting with the last two operators on the lift
of $\phi$, one finds
\begin{align}
    \delta_\sigma\Phi
    & = -\rho(\tfrac1\hbar\,\xi_{\Weyl} + w_{\Weyl})\,\Phi \\
    & = -\tfrac{n-2}2\,\sigma\,\phi
    - \tfrac{n}2\,y^a \nabla_a(\sigma\,\phi)
    - \tfrac{n+2}4\,y^a y^b\,\sigma\,
    \big(\nabla_a \nabla_b - \tfrac16\,R\big)\,\phi \\
    & \quad - \tfrac{n}2\,y^a y^b\,\nabla_a \sigma \nabla_b \phi
    -\tfrac{n-2}6\,y^a y^b\,\phi\,\nabla_a \nabla_b \sigma
    + \tfrac12\,y^2\,\big(\nabla \sigma \cdot \nabla\phi
    + \tfrac16\,\phi\,\Box\sigma\big) + \dots\,,
\end{align}
which leads to
\begin{equation}\label{eq:interm2}
    \rho(F)\,\delta_\sigma\Phi\rvert_{y=0}
    = -\hbar^2\,\tfrac{n+2}2\,\sigma\,\Box\phi
    - \hbar^2\,\tfrac{n-4}6\,\phi\,\Box\sigma
    + \hbar^2\,\tfrac{n(3n-4)+4}{24(n-1)}\,\sigma\,R\,\phi\,.
\end{equation}
Collecting the two terms \eqref{eq:interm1}
and \eqref{eq:interm2}, we finally obtain the action
of a Weyl transformation on the equation of motion,
\begin{equation}
    \delta_\sigma\big(\widehat{f}\phi)
    = -\tfrac{n+2}2\,\sigma\,
    \big(\Box - \tfrac{n-4}{4(n-1)}\,R\big)\,\phi\,,
\end{equation}
as expected: we recover the fact that
the conformal Laplacian sends functions of Weyl weight
$-\tfrac{n-2}2$ to functions of Weyl weight $-\tfrac{n+2}2$.

\section{Feigin--Felder--Shoikhet invariant trace}
\label{app:FFS}
The Hochschild cohomology of the Weyl algebra $\WeylAlg_{2n}$
with values in its linear dual $\WeylAlg^*_{2n}$
is known to be concentrated in degree $2n$ and to be 
one-dimensional \cite{Feigin:1989}.
A representative for this cohomology class,
that we will denote by $\Phi$ hereafter,
was given explicitly by Feigin, Felder and Shoikhet 
\cite{Feigin:2005}, and reads as follows:
\begin{align}
    \Phi(a_0|a_1,\dots,a_{2n}) & = \int_{u\in\Delta_{2n}}
    \exp\Big[\hbar\sum_{0 \leq i<j \leq 2n}
    \big(\tfrac12+u_i-u_j\big)\,\pi_{ij}\Big]\,
    \det\big|\tfrac{\partial}{\partial p^I_a},
    \tfrac{\partial}{\partial y^a_I}\big|_{I=1,\dots,2n} \\
    \nonumber & \hspace{150pt} \times\ 
    a_0(y_0,p_0)\,a_1(y_1,p_1) \dots a_{2n}(y_{2n},p_{2n})\rvert_{y_k=0}\,,
\end{align}
where $\Delta_{2n}$ is the standard $2n$-simplex
which can be defined as
\begin{equation}
    \Delta_{2n} = \big\{(u_1,\dots,u_{2n}) \in \R^{2n}
    \mid u_0 \equiv 0 \leq u_1 \leq u_2 \leq
        \dots \leq u_{2n} \leq 1\big\}\,,
\end{equation}
and 
\begin{equation}
    \pi_{ij} := \frac{\partial}{\partial y^a_i}\,
    \frac{\partial}{\partial p_a^j}
    - \frac{\partial}{\partial p_a^i}\,
    \frac{\partial}{\partial y^a_j}\,,
\end{equation}
and the determinant is taken over the $2n \times 2n$ matrix
whose entries are the operators
$\tfrac{\partial}{\partial p^I_a}$
and $\tfrac{\partial}{\partial y^a_I}$
where the index $I$ runs over $1$ to $2n$,
so that the argument $a_0$ remains unaffected
by this determinant operator.

In practice, we need only the Chevalley--Eilenberg cocycle 
obtained from $\Phi$ by skew-symmetrisation of its arguments,%
\footnote{Recall that the skew-symmetrisation map
is a morphism of complexes between the Hochschild
complex of an associative algebra,
and the Chevalley--Eilenberg of its commutator Lie algebra.}
which we will denote by,
\begin{equation}
    [\Phi](a_0|a_1,\dots,a_{2n})
    = \sum_{\sigma\in\mathcal{S}_{2n}} (-1)^\sigma\,
    \Phi(a_0|a_{\sigma_1},\dots,a_{\sigma_{2n}})\,,
\end{equation}
where $(-1)^\sigma$ denotes the signature
of the permutation $\sigma$. The $n$-cochain defined by
\begin{equation}
    \mu(a_0|a_1,\dots,a_n) := \tfrac1{n!}\,
    \epsilon_{b_1 \dots b_n}\,
    [\Phi](a_0|y^{b_1}, \dots, y^{b_n},a_1,\dots,a_n)\,,
\end{equation}
is {\it almost} a Chevalley--Eilenberg cocycle,
in the sense that it satisfies
\begin{equation}
    \sum_{i=0}^n (-1)^i\,
        \mu\big([a_{-1},a_i]_\ast | a_0, \dots,a_n\big)
    + \sum_{0 \leq i < j \leq n} (-1)^{i+j}\,
        \mu\big(a_{-1} | [a_i, a_j]_\ast, a_0,\dots,a_n\big)
    = \tfrac{\partial}{\partial p_a}\,
        \varphi_a(a_{-1}|a_0, \dots, a_n)\,,
\end{equation}
where
\begin{equation}
    \varphi_a(a_{-1} | a_0, \dots, a_n)
    = \tfrac1{(n-1)!}\,\epsilon_{a b_1\dots b_{n-1}}\,
    [\Phi](a_{-1}|y^{b_1}, \dots, y^{b_{n-1}},a_0,\dots,a_n)\,,
\end{equation}
i.e. it verifies the cocycle condition modulo
a total derivative in $p$. As a first step towards
simplifying the expression of $\mu$, let us note that
\begin{equation}
    \det\big|\tfrac{\partial}{\partial y^a_I},
    \tfrac{\partial}{\partial p_a^I}\big|
    = \sum_{\sigma \in \mathcal{S}_{n|n}}
    (-1)^\sigma\,\epsilon^{a_1 \dots a_n}\,
    \frac{\partial}{\partial y^{a_1}_{\sigma_1}}
    \dots \frac{\partial}{\partial y^{a_n}_{\sigma_n}}\,
    \epsilon_{b_1 \dots b_n}\,
    \frac{\partial}{\partial p_{b_1}^{\sigma_{n+1}}}
    \dots \frac{\partial}{\partial p_{b_n}^{\sigma_{2n}}}\,
\end{equation}
where $\mathcal{S}_{n|n}$ denotes the set of permutations
of $2n$ elements which preserve the order of the first $n$
and the last $n$ elements separately,
i.e. $\sigma_1 < \sigma_2 < \dots < \sigma_n$
and $\sigma_{n+1} < \sigma_{n+2} < \dots < \sigma_{2n}$, and
\begin{align}
    \tfrac1{n!}\,\epsilon_{a_1 \dots a_n}\,&
    \sum_{\sigma\in\mathcal{S}_{2n}} (-1)^\sigma\,
    y^{a_1}_{\sigma_1} \dots y^{a_n}_{\sigma_n}\,
    a_1(y_{\sigma_{n+1}}, p_{\sigma_{n+1}})
    \dots a_n(y_{\sigma_{2n}},p_{\sigma_{2n}})\\
    & = \epsilon_{a_1 \dots a_n}\,
    \sum_{\sigma\in\mathcal{S}_{n|n}} (-1)^\sigma\,
    y^{a_1}_{\sigma_1} \dots y^{a_n}_{\sigma_n}\,
    \sum_{\tau\in\mathcal{S}_n} (-1)^\tau\,
    a_{\tau_1}(y_{\sigma_{n+1}}, p_{\sigma_{n+1}})
    \dots a_{\tau_n}(y_{\sigma_{2n}},p_{\sigma_{2n}})\,,
    \nonumber
\end{align}
so that, put together, these two formulae yield
\begin{align}
    \det\big|\tfrac{\partial}{\partial y^a_I},
    \tfrac{\partial}{\partial p_a^I}\big|
    & \Big(\tfrac1{n!}\,\epsilon_{a_1 \dots a_n}\,
    \sum_{\sigma\in\mathcal{S}_{2n}} (-1)^\sigma\,
    y^{a_1}_{\sigma_1} \dots y^{a_n}_{\sigma_n}
    a_1(y_{\sigma_{n+1}}, p_{\sigma_{n+1}})
    \dots a_n(y_{\sigma_{2n}},p_{\sigma_{2n}})\Big) \\
    & = (2n)!\,\sum_{\substack{\{i_1<\dots<i_n\}\\
                    \subset \{1,\dots,2n\}}}\
    \sum_{\sigma\in\mathcal{S}_n} (-1)^\sigma\,
    \epsilon_{a_1 \dots a_n}\,
    \frac{\partial a_{\sigma_1}}{\partial p_{a_1}}
    (y_{i_1},p_{i_1}) 
    \dots \frac{\partial a_{\sigma_n}}{\partial p_{a_n}}
     (y_{i_n},p_{i_n})\,,
\end{align}
where the first sum is taken over all \emph{ordered} subsets
of $n$ integers in the set $\{1,\dots,2n\}$.
We are now in position of writing down the cochain $\mu$:
for any $a_0,a_1,\dots,a_n \in \WeylAlg_{2n}$,
it is given explicitly by 
\begin{equation}
    \mu(a_0|a_1,\dots,a_n)
    = (2n)!\,\sum_{\sigma\in\mathcal{S}_n} (-1)^\sigma\,
    \int_{u\in\Delta_{2n}} \mathscr{D}(u;a_0,a_{\sigma_1},
    \dots,a_{\sigma_n})\rvert_{y=0}\,,
\end{equation}
where
\begin{equation}
    \mathscr{D}(u;-) = \sum_{f\in\Delta([n],[2n])}\
    \exp\Big[\hbar\sum_{0 \leq i<j \leq n}
    \big(\tfrac12+u_{f(i)}-u_{f(j)}\big)\,\pi_{ij}\Big]
    \epsilon_{a_1 \dots a_n}\,\big(1 \otimes
    \tfrac{\partial}{\partial p_{a_1}} \otimes
    \dots \otimes \tfrac{\partial}{\partial p_{a_n}}\big)\,,
\end{equation}
and 
\begin{equation}
    \Delta([k],[l]) := \big\{f: \{1,2,\dots,k\}
    \longrightarrow \{1,2,\dots,l\} \mid f(i) < f(j)\,,\ 
    1 \leq i < j \leq k\big\}
\end{equation}
denotes the set of order-preserving maps from the set
$[k]$ of the first $k$ integers, to the set $[l]$
of the first $l$ integers. Note that by convention,
we put $f(0)=0$ and $u_0=0$.

\paragraph{Trace on the deformed algebra of functions.}
Suppose that $a_1, \dots, a_n$ are linear in $p$,
and write $\tfrac{\partial a}{\partial p_b} = a^b(y)$
for their derivative with respect to $p$.
Then the above operator collapses to
\begin{equation}
    \mathscr{D}(u;a_0,a_1,\cdots,a_n)
    = \sum_{f\in\Delta([n],[2n])} \exp\Big[\hbar\sum_{i=1}^n
    \big(u_{f(i)}-\tfrac12\big)\,\tfrac{\partial}{\partial p_a}\,
    \tfrac{\partial}{\partial y_i^a}\Big]\,a_0(y,p)
    \times \epsilon_{b_1 \dots b_n}\,
    a^{b_1}_1(y_1) \dots a^{b_n}_n(y_n)\,,
\end{equation}
thereby exhibiting a clear distinction between
the arguments: the zeroth one will only receive
derivative with respect to $p$, while the remaining
$n$ arguments will only receive derivatives
with respect to $y$.
Now consider the case where $a_0=F(y,p)$,
and all other arguments are equal to the Fedosov connection,
$a_1=\dots=a_n=A$.
Since $A$ is linear in $p$ we can write it as
\begin{equation}
    A(y,p) = \dR x^\mu\,e_\mu^a\,A_a{}^b(y)\,p_b\,,
\end{equation}
and introducing the notation
\begin{equation}\label{eq:alt}
    \mathbb{A}(y_1, \dots, y_n)
    := \epsilon^{a_1 \dots a_n}\,\epsilon_{b_1 \dots b_n}\,
    A_{a_1}{}^{b_1}(y_1) \cdots A_{a_n}{}^{b_n}(y_n)\,,
\end{equation}
we end up with
\begin{equation}
    \mathscr{D}(u;F,A,\dots,A) = \dR^nx\,|e|\,
    \sum_{f\in\Delta([n],[2n])} \exp\Big[\hbar\sum_{i=1}^n
    \big(u_{f(i)}-\tfrac12\big)\,\tfrac{\partial}{\partial p_a}\,
    \tfrac{\partial}{\partial y_i^a}\Big]\,
    F(y,p) \times \mathbb{A}(y_1, \dots, y_n)\,,
\end{equation}
where $|e|$ is the determinant of the vielbein.
This formula exhibits a couple of properties:
\begin{itemize}
\item First, as we noticed earlier, the argument $F(y,p)$
is the only one to receive derivatives with respect to $p$.
This means that in order to compute $\mu(F|A,\dots,A)$,
one only needs to know $F\rvert_{y=0}$, the $y$-independent
part of the symbol $F$.

\item Second, the integral over the simplex will produce
some combinatorial coefficients
\begin{equation}
    \sum_{f\in\Delta([n],[2n])} \int_{\Delta_{2n}}
    \big(u_{f(\ell_1)}-\tfrac12\big)\,\cdots\,
    \big(u_{f(\ell_k)}-\tfrac12\big)\,,
\end{equation}
which depends on a $k$-tuple of integers
$(\ell_1, \dots, \ell_k)$ comprised between $1$ and $n$.
In fact, one can refine this dependency a little bit
by remarking that if two $k$-tuple are related by
a permutation $\tau \in \mathcal{S}_k$, the associated
coefficients are equal,
so that these coefficients may as well be labeled
by partitions of $k$.
\end{itemize}

Putting this together, one ends up with
\begin{align}
    \mu(F|A,\dots,A) & = \dR^nx\,|e|\,
    \sum_{k\geq0} \mu^{\scriptscriptstyle\nabla}_{a_1 \dots a_k}(x)
    \frac{\partial^k}{\partial p_{a_1} \cdots \partial p_{a_k}}
    F(y,p)\big|_{y=0}\,,
\end{align}
where $\mu^{\scriptscriptstyle\nabla}_{a_1 \dots a_k}(x)$
are polynomials in the (covariant derivatives of the) curvature
of $\nabla$ which is obtained by computing the term
of order $\hbar^k$ in
\begin{equation}
    \sum_{f\in\Delta([n],[2n])} \int_{\Delta_{2n}}
    \exp\Big[\hbar\sum_{i=1}^n \big(u_{f(i)}-\tfrac12\big)\,
    \tfrac{\partial}{\partial y_i^a}
    \otimes \tfrac{\partial}{\partial p_a}\Big]\,
    \mathbb{A}(y_1, \dots, y_n) \otimes F(y,p)\big|_{y_i=0}\,.
\end{equation}

\section{Curvature expansion}
\label{app:curvature}
Since the components of $\Completion$ are constructed
from the curvature tensor of $\nabla$, its covariant
derivatives and contractions thereof, we can re-arrange
its expansion in power of the curvature, which appears through 
\begin{equation}
    \Curvature \equiv -\tfrac13\,\dR x^\mu\,R_{\mu a}{}^c{}_b\,
    y^a y^b\,p_c\,,
\end{equation}
namely we write $\Completion=\sum_{k\geq1} \Completion^{(k)}$ 
where $\Completion^{(k)}$ is of order $k$ in $\Curvature$
and its derivatives. Let us now evaluate
its defining equation \eqref{eq:rec_A}
at order $n$ in $\Curvature$, 
\begin{equation}
    \Completion^{(k)} = \Curvature\,\delta_{k,1}
    + \SymDer \Completion^{(k)}
    + \sum_{l=1}^{k-1} \tfrac{1}{2\hbar}\,
    h\big[\Completion^{(l)}, \Completion^{(k-l)}\big]_\ast\,,
\end{equation}
which we can re-write, for $k>1$, as
\begin{equation}\label{eq:rec_curv}
    \Completion^{(k)} = \tfrac1{2\hbar}\,
    \sum_{l=1}^{k-1} \Taylor h\,
    \big[\Completion^{(l)}, \Completion^{(k-l)}\big]_\ast\,,
    \qquad\text{with}\qquad 
    \Taylor := \frac1{1-\SymDer} = \sum_{m=0}^\infty \SymDer^m\,.
\end{equation}
The first few orders in this curvature expansion
\begin{equation}
    \Completion^{(1)} = \Taylor\Curvature\,,
    \qquad 
    \Completion^{(2)} = \tfrac1{2\hbar}\,\Taylor
        \Bracket{\Curvature, \Curvature}\,,
    \qquad 
    \Completion^{(3)}
    = \tfrac1{2\hbar^2}\,\Taylor\Bracket{\Curvature,
        \Bracket{\Curvature,\Curvature}}\,,
\end{equation}
\begin{equation}
    \Completion^{(4)} = \tfrac1{2\hbar^3}\,\Taylor
    \Bracket{\Curvature,\Bracket{\Curvature,
        \Bracket{\Curvature,\Curvature}}}
    +\tfrac1{8\hbar^3}\,\Taylor
        \Bracket{\Bracket{\Curvature,\Curvature},
            \Bracket{\Curvature,\Curvature}}\,,
\end{equation}
where we introduce the bracket
\begin{equation}
    \Bracket{-,-}
        := h\,\big[\Taylor(-),\Taylor(-)\big]_\ast\,,
\end{equation}
as a shorthand notation. This approach has a couple
of advantages: first, it allows us to access in one go
whole pieces of $\Completion$ at arbitrary order in $y$,
and second, the recursion in order of curvature
exhibits an interesting structure, namely it appears
that it is controlled by the grafting (non-planar)
binary trees. Indeed, denoting the operator $\Taylor$
by an edge, and the composition of the star-commutator
with the contracting homotopy by a vertex, i.e.
\begin{equation}
    \Taylor(X) = 
    \begin{tikzpicture}[baseline=(m1), scale=0.75]
        \node[above] (in) at (0,0.5) {$X$};
        \node (m1) at (0,0) {};
        \node (out) at (0,-0.5) {};
        \draw[thick] (in) -- (out);
    \end{tikzpicture}
    \hspace{50pt}
    \tfrac1{2\hbar}\,h[X, Y]_\ast = 
    \begin{tikzpicture}[baseline=(m2), scale=0.75]
        \node[above] (1st) at (-1,0.5) {$X$};
        \node[above] (2nd) at (1,0.5) {$Y$};
        \node (out) at (0,-1) {};
        \node[vertex] (m2) at (0,0) {};
        \draw[dashed] (1st) -- (m2);
        \draw[dashed] (2nd) -- (m2);
        \draw[dashed] (m2) -- (out);
    \end{tikzpicture}
\end{equation}
where the diagrams should be read from top to bottom,
so that for instance
\begin{equation}
    \Taylor\Bracket{X,Y}\ =
    \begin{tikzpicture}[baseline=(m2), scale=0.75]
        \node[above] (1st) at (-1,0.5) {$X$};
        \node[above] (2nd) at (1,0.5) {$Y$};
        \node (out) at (0,-0.75) {};
        \node[vertex] (m2) at (0,0) {};
        \draw[thick] (1st) -- (m2) -- (out);
        \draw[thick] (2nd) -- (m2);
    \end{tikzpicture}
\end{equation}
one can re-write the recursion relation \eqref{eq:rec_curv} as
\begin{equation}
    \Completion^{(k)} = \sum_{l=1}^{k-1}\
    \begin{tikzpicture}[baseline=(m2), scale=0.75]
        \node[above] (1st) at (-1,0.5) {$\Completion^{(l)}$};
        \node[above] (2nd) at (1,0.5) {$\Completion^{(k-l)}$};
        \node (out) at (0,-0.75) {};
        \node[vertex] (m2) at (0,0) {};
        \draw[dashed] (1st) -- (m2);
        \draw[dashed] (2nd) -- (m2);
        \draw[thick] (m2) -- (out);
    \end{tikzpicture}
\end{equation}
for all $k > 1$. Now it becomes relatively easy
to see that the result of this recursion relation
is to express $\Completion^{(k)}$ as a sum 
over all \emph{rooted planar binary trees with $k$ leaves.}
Indeed, the latter can be obtained by successively
\emph{grafting}---meaning summing over all trees
resulting from attaching the root of a tree to the leaves
of another one---the rooted binary tree with a single vertex
to itself, $k$ times, which is exactly what the above relation
produces. Taking into account the fact that $\Bracket{-,-}$
is antisymmetric amounts to identifying any two rooted
planar binary trees which can be related by permuting
the two leaves at each node, that is, one should 
sum over rooted \emph{non-planar} binary trees
and take into account the number of planar ones
that it is equivalent to as a multiplicity.

Having worked out the recursion formula for $\Completion$
in order of the curvature, we can do the same
for the lift of symbols. Indeed, resumming the defining
relation \eqref{eq:rec_F} for the lift $F(x;y,p)$
of a symbol $f(x,p)$ yields,
\begin{equation}
    F = \SymDer F + \tfrac1\hbar\,h\,[\Completion,F]_\ast
\end{equation}
from which we can extract the order $k$ piece via
\begin{equation}
    F^{(0)} = \Taylor f\,,
    \qquad 
    F^{(k>0)} = \tfrac1\hbar\,\sum_{l=0}^{k-1} \Taylor h\,
    [\Completion^{(k-l)}, F^{(l)}]_\ast\,.
\end{equation}
The first few orders are given by
\begin{equation}
    F^{(1)} = \tfrac1\hbar\,\Taylor\,\Bracket{\Curvature,f}\,,
    \qquad 
    F^{(2)} = \tfrac1{2\hbar^2}\,\Taylor
    \Big(\Bracket{\Bracket{\Curvature,\Curvature},f}
    + 2\,\Bracket{\Curvature,\Bracket{\Curvature,f}}\Big)\,,
\end{equation}
and
\begin{equation}
    F^{(3)} = \tfrac1{2\hbar^3}\,\Taylor
    \Big(\Bracket{\Bracket{\Curvature,
            \Bracket{\Curvature,\Curvature}},f}
    + \Bracket{\Bracket{\Curvature,
        \Curvature},\Bracket{\Curvature,f}}
    + \Bracket{\Curvature,
        \Bracket{\Bracket{\Curvature,\Curvature},f}}
    + 2\,\Bracket{\Curvature,
        \Bracket{\Curvature,\Bracket{\Curvature,f}}}\Big)\,.
\end{equation}

Finally, the same can be done for the lift of a function
$\phi(x)$ to a covariantly constant section $\Phi(x;y)$
of the Fock bundle: the re-summed for of the recursion
relation \eqref{eq:rec_phi} reads
\begin{equation}
    \Phi = \SymDer\Phi
    + \tfrac1\hbar\,h\,\rho(\Completion)\Phi\,.
\end{equation}
which, when evaluated at order $n$ in $\Curvature$ 
gives us
\begin{equation}
    \Phi^{(0)} = \Taylor\,\phi\,,
    \qquad 
    \Phi^{(k>0)} = \tfrac1\hbar\,\sum_{l=0}^{k-1} \Taylor h\,
    \rho(\Completion^{(k-l)})\Phi^{(l)}\,.
\end{equation}
The first few orders read
\begin{equation}
    \Phi^{(1)} = \tfrac1\hbar\,\Taylor\big(\Rep{\Curvature}{\phi}\big)\,,
    \qquad 
    \Phi^{(2)} = \tfrac1{2\hbar^2}\,\Taylor
    \Big(\Rep{\Bracket{\Curvature,\Curvature}}{\phi}
    + 2\,\Rep{\Curvature}{\big(\Rep{\Curvature}{\phi}\big)}\Big)\,,
\end{equation}
and
\begin{equation}
    \Phi^{(3)} = \tfrac1{2\hbar^3}\,\Taylor
    \Big(\Rep{\Bracket{\Curvature,
            \Bracket{\Curvature,\Curvature}}}{\phi}
    + \Rep{\Bracket{\Curvature,
        \Curvature}}{(\Rep{\Curvature}{\phi})}
    + \Rep{\Curvature}{\big(\Rep{\Bracket{\Curvature,\Curvature}}{\phi}\big)}
    + 2\,\Rep{\Curvature}{\big(\Rep{\Curvature}{(\Rep{\Curvature}{\phi})}\big)}\Big)\,,
\end{equation}
where we introduced the shorthand notation
\begin{equation}
    \Rep{\bullet}{(-)}
    := h\,\rho\big(\Taylor(\bullet)\big)\,\Taylor(-)\,,
\end{equation}
for the sakes of conciseness. The term of order $k$
will be almost identical to that of $F$,
except for the replacement of $f$ with $\phi$,
and any bracket of the form $\Bracket{-,f}$,
i.e. whose second argument is $f$, with $\Rep{(-)}{\phi}$.
This is of course not surprising since the only 
difference between the two case is the representation
of the Weyl algebra in which the covariantly constant
section that we are solving for sits in: the adjoint
for symbols like $f$ and the Fock one for functions like $\phi$.

\paragraph{Simplifying the elementary operations.}
Let us try to find a concise expression for the operator $\Taylor$.
To do so, first notice that
\begin{equation}
    \SymDer\alpha = h\nabla\alpha = \tfrac1N\,
    \big(y^a \nabla_a \alpha - \nabla\,N\,h\alpha\big)\,,
\end{equation}
so that, on forms valued in the Weyl algebra
which are \emph{annihilated} by the homotopy operator $h$,
one finds
\begin{equation}
    h\alpha=0
    \qquad\Longrightarrow\qquad 
    \SymDer\alpha = \tfrac1N\,y \cdot \nabla\alpha
    = y \cdot \nabla\,\tfrac1{N+1}\,\alpha\,,
\end{equation}
where we used the fact that $y \cdot \nabla$ is obviously
of degree $1$ in $y$, and hence increases the eigenvalue
of the number operator $N$, a fact we should take into account
when moving the latter to the right of the former.
Repeating this operation, we end up with
\begin{equation}
    \Taylor = \sum_{k\geq0} (y \cdot \nabla)^k\,\tfrac1{(N+1)_k}
    = \sum_{k\geq0} \tfrac1{k!}\,(y \cdot \nabla)^k\,
    \tfrac{(1)_k}{(N+1)_k}
    \equiv {}_1F_1\big[1;N+1; y \cdot \nabla\big]\,,
\end{equation}
where $(a)_k=a(a+1)\dots(a+k-1)$ is the raising Pochhammer symbol,
and where we used the fact that $(1)_k=k!$ to recognize
the \emph{confluent hypergeometric function}.
Let us stress that the above expression is valid only
for $\Taylor$ acting on elements in ${\rm Ker}(h)$,
which is enough for us since we are interested in applying it 
to either $\Completion$ or a $0$-form, both annihilated by $h$,
by definition.

We can now use the integral representation of the confluent
hypergeometric function,
\begin{equation}\label{eq:1F1}
    {}_1F_1\big[a;b;x\big] := \sum_{k\geq0} \tfrac{x^k}{k!}\,
    \tfrac{(a)_k}{(b)_k} = \frac{\Gamma(b)}{\Gamma(a)\,\Gamma(b-a)}\,
    \int_0^1 \dR t\,e^{t\,x}\,t^{a-1}\,(1-t)^{b-a-1}\,,
\end{equation}
which holds for $\Re(b) > \Re(a) > 0$. In our case,
both parameters are positive integer, and verify the inequality
except if $N=0$, i.e. if we act on a $y$-independent $0$-form.
This case is particularly simple to treat since the hypergeometric
series collapses to an ordinary exponential, i.e.
\begin{equation}\label{eq:j_0-form}
    \Taylor\rvert_{\Functions(T^*\Base)} = e^{y \cdot \nabla}\,.
\end{equation}
We can therefore exclude this case, that is consider $N>0$,
and use the above integral representation to re-write $\Taylor$ as
\begin{equation}
    \Taylor = N\,\int_0^1 \dR t\,e^{t\,y \cdot \nabla}\,(1-t)^{N-1}
    =\int_0^1\, \dR t\,e^{(1-t)\,y \cdot \nabla}\,
    \tfrac{\dR}{\dR t}\,t^N\,,
\end{equation}
where we used the change of variable $t \to 1-t$
before recognizing the derivative. We therefore find
\begin{align}
    \Taylor(\alpha) & = \int_0^1 \dR t\,e^{(1-t)\,y \cdot \nabla}\,
    \tfrac{\dR}{\dR t}\alpha(x,t\,\dR x, t\,y,p) \\
    & = \alpha + y \cdot \nabla\,\int_0^1 \dR t\,
    e^{(1-t)\,y \cdot \nabla}\,\alpha(x,t\,\dR x, t\,y,p)\,,
\end{align}
for any Weyl algebra-valued form $\alpha$ such that $h\alpha=0$.
Remark that this last form obtained by integrating by part, namely
\begin{equation}
    \Taylor = 1 - \int_0^1 \dR t\,\tfrac{\dR}{\dR t}
    \big(e^{(1-t)\,y\cdot\nabla}\big)\,t^N\,,
\end{equation}
also makes sense for $N=0$, since it reproduces \eqref{eq:j_0-form}.

\paragraph{Lifts at first order in curvature.}
Let us now use this operator to compute the lift 
of symbols and wave functions at first order in curvature.
Starting with the latter, we need to first compute
\begin{align}
    \rho(\Taylor \Curvature)\,\Taylor \phi 
    & = \exp\Big(\!-\hbar\,\partial_p \cdot
    \big[\tfrac12\,\partial_{y_1} + \partial_{y_2}\big]\Big)
    \int_0^1 \dR u\,u^2\,e^{(1-u)\,y_1 \cdot \nabla}\,
    \big(-\dR x^\mu\,R_{\mu a}{}^c{}_b\,y_1^a y_1^b\,p_c\big)\,
    \times\, e^{y_2 \cdot \nabla} \phi\big|_{p=0, y_1=y=y_2} 
    \nonumber \\
    & = -\tfrac\hbar2\,\int_0^1 \dR u\,u^2\,
    e^{(1-u)\,y \cdot \nabla}\,
    \big(\dR x^\mu\,R_{\mu a}\,y^a\big)
    \times e^{y \cdot \nabla} \phi + (\cdots)\,,
\end{align}
where the dots denote terms annihilated by the contracting
homotopy $h$. Applying the latter composed with $\Taylor$,
we end up with
\begin{align}
    \Phi^{(1)} & \equiv \tfrac1\hbar\,
    \Taylor\,h\,\rho(\Taylor \Curvature)\,
    \Taylor \phi = -\tfrac12\,\int_{[0,1]^3} \dR t\,
    \dR s\,\dR u\ e^{(1-t)\,y \cdot [\nabla_1+\nabla_2]} \\
    & \hspace{150pt} \times\ \tfrac{\dR}{\dR t}\,\Big(s\,t^2\,u^2\,
    e^{t\,s\,y \cdot [\nabla_1 + (1-u)\,\nabla_2]}
    \phi(x_1)\,R_{ab}(x_2)\,y^a y^b\Big)\rvert_{x_1=x=x_2}\,,
    \nonumber
\end{align}
where $\nabla_i$ denotes the covariant derivative
with respect to $x_i$. Evaluating this formula up
to third order in $y$ yields
\begin{equation}
    \Phi = \phi + y^a\,\nabla_a\phi + \tfrac12\,y^a y^b\,
    \big(\nabla_a \nabla_b - \tfrac16\,R_{ab}\big)\phi
    + \tfrac16\,y^a y^b y^c\,\big(\nabla_a \nabla_b \nabla_c
    -\tfrac12\,R_{ab}\,\nabla_c -\tfrac14\,\nabla_a R_{bc}\big)\phi
    + (\dots)\,,
\end{equation}
as previously derived from the defining recursion relation
for the lift of the wave function $\phi$ in the Fock bundle.
We can re-write this lift up to first order in curvature as
\begin{equation}
    \Phi(x;y) = \Big(\tau^{(0)}(y \cdot \nabla_1)
    + \tau^{(1)}(y \cdot \nabla_1, y \cdot \nabla_2)\,
    \tfrac12\,y^a y^b\,R_{ab}(x_2)
    + \dots\Big)\phi(x_1)\rvert_{x_i=x}\,,
\end{equation}
with $\tau^{(0)}(z_1) = e^{z_1}$ and
\begin{equation}
    \tau^{(1)}(z_1,z_2)
    = -\int_{[0,1]^3} \dR t\,\dR s\,\dR u\
    e^{(1-t)\,[z_1+z_2]}\ \tfrac{\dR}{\dR t}\,\Big(s\,t^2\,u^2\,
    e^{t\,s\,[z_1 + (1-u)\,z_2]}\Big) 
    = -\tfrac16\,e^{z_1}\,{}_1F_1\big[2;4;z_2\big] 
\end{equation}
which, upon using the integral representation \eqref{eq:1F1},
can be expressed as
\begin{equation}
    \tau^{(1)}(z_1,z_2) = e^{z_1}\,
    \int_0^1 \dR t\ (t-1)\,t\,e^{t\,z_2}\,.
\end{equation}
and hence
\begin{equation}\label{eq:Phi1}
    \Phi(x;y) = \Big(1 +\tfrac12\,y^a y^b\,
    \int_0^1 \dR t\ t\,(t-1)\,e^{t\,y \cdot \nabla} R_{ab} 
    + \dots\Big)\,e^{y \cdot \nabla} \phi\,.
\end{equation}

Now let us turn our attention to the piece of first order
in curvature of the lift of $f$, for which we need to compute
\begin{align}
    [\Taylor \Curvature, \Taylor f]_\ast
    & = \sum_{\sigma=\pm} \sigma\,
    \exp\Big(\tfrac{\sigma\hbar}{2}\,
    \big[\partial_{y_1} \cdot \partial_{p_2}
    - \partial_{p_1} \cdot \partial_{y_2}\big]\Big) \\
    & \hspace{50pt} \times \,\int_0^1 \dR u\,
    e^{(1-u)\,y_1 \cdot \nabla}\,\big(-u^2\,\dR x^\mu\,
    R_{\mu\,a}{}^c{}_b y^a_1 y^b_1\,p_{1\,c}\big)
    \times e^{y_2 \cdot \nabla} f(p_2)
    \big|_{\substack{y_1=y=y_2 \\ p_1=p=p_2}} \\
    & = \sum_{\sigma=\pm} \sigma\,
    \exp\Big(\tfrac{\sigma\hbar}{2}\,
    \partial_{y_1} \cdot \partial_{p_2}\Big)
    \int_0^1 \dR u\,e^{(1-u)\,y_1 \cdot \nabla} 
    \big(u^2\,\dR x^\mu\,R_{\mu\,a}{}^c{}_b\,y_1^a y_1^b\big) \\
    & \hspace{180pt} \times\,
    \big(-p_{1\,c} + \tfrac{\sigma\hbar}{2}\,\nabla_c\big)\,
    e^{y_2 \cdot \nabla}\,f(p_2)
    \big|_{\substack{y_1=y=y_2\\p_1=p=p_2}} \\
    & = \tfrac{\hbar}{2}\,\sum_{\sigma=\pm} 
    \int_0^1 \dR u\ e^{(1-u)
    [\,y + \frac{\sigma\hbar}{2}\,\partial_p] \cdot \nabla}
    \Big(u^2\,\dR x^\mu\,R_{\mu\,a}{}^c{}_b\,
    \big[y^b\,\tfrac{\partial}{\partial p_a}
    + \tfrac{\sigma\hbar}{2}\,
    \tfrac{\partial^2}{\partial p_a \partial p_b}\big]\Big) \\
    & \hspace{150pt} \times\,
    \big(-p_c' + \tfrac{\sigma\hbar}{2}\,\nabla_c\big)\,
    e^{y \cdot \nabla}\,f(p)\rvert_{p'=p} + (\dots)\,, 
\end{align}
where the dots denote terms that are annihilated by $h$.
Applying it followed by $\Taylor$, one finds
\begin{align}
    F^{(1)} & \equiv \tfrac1\hbar\,\Taylor h\,
    [\Taylor\Curvature, \Taylor f]_\ast 
    = \tfrac12\,\sum_{\sigma=\pm} 
    \int_{[0,1]^3} \dR s\,\dR t\,\dR u\
    e^{(1-t)\,y \cdot [\nabla_1 + \nabla_2]} \\
    & \hspace{150pt} \times\ \tfrac{\dR}{\dR t}\,
    \Big(t\,u^2\,e^{s\,t\,y \cdot [(1-u)\,\nabla_1+\nabla_2]
    +\frac{\sigma\hbar}2\,(1-u)\,\partial_p \cdot \nabla_1}\,
    y^d\,R_{da}{}^c{}_b(x_1) \\ & \hspace{150pt}
    \times\,\big[t\,s\,y^b\,\tfrac{\partial}{\partial p_a}
    + \tfrac{\sigma\hbar}{2}\,
    \tfrac{\partial^2}{\partial p_a \partial p_b}\big]\,
    \big(-p_c' + \tfrac{\sigma\hbar}{2}\,\nabla_c\big)\,
    f(x_2,p)\Big)\big|_{\substack{x_1=x=x_2\\p'=p}}\,.
\end{align}
After explicitly performing the integrals as for the lift
of $\phi$, one finds 
\begin{align}
    F^{(1)}(x,p;y) & = \tfrac12\,e^{y \cdot \nabla_1}\,
    \sum_{\substack{k,l\geq0\\\sigma=\pm}}
    \big(\tfrac{\sigma\hbar}{2}\big)^k\,
    \tfrac{(\partial_p \cdot \nabla_2)^k}{k!}\,
    \tfrac{(y \cdot \nabla_2)^l}{l!}\,y^a\,R_{da}{}^c{}_b(x_2)\,
    \tfrac{1}{(k+1)(l+k+2)(l+k+3)} \\
    & \hspace{100pt} \times\,
    \Big[y^b\,\tfrac{\partial}{\partial p_d}
    + \tfrac{\sigma\hbar}{2}\,\tfrac{l+2k+4}{(k+2)}\,
    \tfrac{\partial^2}{\partial p_b \partial p_d}\Big]\,
    \big(p_c' - \tfrac{\sigma\hbar}{2}\,\nabla_c\big)\,
    f(x_1,p)\big|_{\substack{x_i=x,\\p'=p}} \\
    & = \tfrac12\,\sum_{\sigma=\pm}\int_{[0,1]^2} \dR s\,\dR t\
    t\,(1-t)\,e^{y \cdot \nabla_1
    + t\,[y + s\frac{\sigma\hbar}2\,\partial_p] \cdot \nabla_2}\,
    \big[p' - \tfrac{\sigma\hbar}2\,\nabla_1\big]_c 
    \label{eq:F1} \\
    & \hspace{50pt} \times\,y^a\,\Big(y^b
    + \sigma\hbar\,\big[1 + \tfrac{t}2\,(1-s)\,
    y \cdot \nabla_2\big]\,\tfrac{\partial}{\partial p_b}\Big)\,
    R_{da}{}^c{}_b(x_2)\,\tfrac{\partial}{\partial p_d}
    f(x_1,p)\big|_{\substack{x_i=x,\\p'=p}}\,, 
\end{align}
for the piece of first order in curvature of the lift
$\tau(f)$ of any symbol $f(x,p)$.

\paragraph{Generating function.}
Combining the first order lift $\Phi^{(1)}$
of the scalar field given in \eqref{eq:Phi1} with the fact that 
\begin{equation}
    \rho(\Taylor f)\rvert_{y=0}
    = \exp\Big(\!-\hbar\partial_p \cdot \big[\tfrac12\,\nabla
    + \partial_y\big]\Big)f\,\big|_{y=0=p}
\end{equation}
for a symbol $f=f(x,p)$, i.e. $y$-independent,
one ends up with
\begin{align}
    \rho(F^{(0)})\Phi^{(1)}\rvert_{y=0}
    & = \tfrac{\hbar^2}{2}\,\int_0^1 \dR t\ (t-1)t\,
    \times\, e^{-\hbar\,\partial_p \cdot [\nabla_1 + t\,\nabla_2
    +\frac12\,\nabla_3]}\,\phi(x_1)\,R_{ab}(x_2)\,
    \tfrac{\partial^2}{\partial p_a \partial p_b} f(x_3,p)
    \big|_{x_i=x, p=0}\,,
\end{align}
upon using the BCH formula. Note that the commutator appearing
in these manipulations can be discarded as a consequence
of the fact that we are working at first order in curvature.
Now applying the quantization map to the first order lift
$F^{(1)}$ of a symbol given in \eqref{eq:F1},
we end up with,
\begin{align}
    \rho(F^{(1)})\Phi^{(0)}\rvert_{y=0} & = \tfrac{\hbar^2}{8}\,
    \sum_{\sigma=\pm}\int_{[0,1]^2} \dR s\,\dR t\ t(t-1)\,
    e^{-\frac{\hbar}2\,\partial_p \cdot
    [2\nabla_1 + t(1-\sigma\,s)\,\nabla_2+\nabla_3]} \\
    & \hspace{50pt} \times\,
    \Big[(1-2\sigma) + \tfrac{\sigma\hbar}2\,t(1-s)\,
    \partial_p \cdot \nabla_2\Big]\,\phi(x_1)\,R_{ab}(x_2)\,
    \tfrac{\partial^2}{\partial p_a \partial p_b}
    f(x_3,p)\big|_{\substack{x_i=x,\\p=0}}\,, \nonumber
\end{align}
after using the BCH formula a few times as before.
Performing the integrals, multiplying the result by $\phi^*$,
and integrating by part so that all derivatives on $f$
are re-distributed on $\phi$, $\phi^*$ and the curvature,
one ends up with
\begin{equation}
    {\cal J}(x|u) = e^{-\frac\hbar2\,u \cdot[\nabla_1-\nabla_2]}\,
    \Big(1 - \tfrac{\hbar^2}8\,
    {\rm sinhc}(\tfrac\hbar4\,u \cdot \nabla_3)\,
    R_{ab}(x_3)\,u^a u^b + {\cal O}(R^2)\Big)\,\phi(x_1)\,\phi^\dagger(x_2)\big|_{x_i=x}\,,
\end{equation}
where
\begin{equation}
    {\cal J}(x|u) := \sum_{s \geq 0} \tfrac{(-\hbar)^s}{2^s\,s!}\,
    J_{a_1 \dots a_s}(x)\,u^{a_1} \dots u^{a_s}\,,
\end{equation}
is the generating function for the higher spin currents, and 
\begin{equation}
    {\rm sinhc}(z) := \frac{\sinh(z)}{z}\,.
\end{equation}
is the hyperbolic version of the $\rm sinc$ function.
 
\newpage
\providecommand{\href}[2]{#2}\begingroup\raggedright\endgroup

\end{document}